\documentclass[manuscript]{aastex}

\usepackage{amsmath,hyperref}

\newcommand{\fflex}{{\mathcal F}}
\newcommand{\gflex}{{\mathcal G}}

\def\sersic{\mbox{S\'ersic}}

\slugcomment{}

\shorttitle{Measuring Flexion in A1689 with AIM}
\shortauthors{Cain et al.}

\begin{document}

\title{Measuring Gravitational Lensing Flexion in Abell 1689 Using an Analytic Image Model}

\author{Benjamin Cain}
\affil{MIT Kavli Institue for Astrophysics and Space Research/\\University of California Davis Department of Physics\\One Shields Avenue, Davis, CA 95616}
\email{bmcain@ucdavis.edu}

\author{Paul L.~Schechter}
\affil{MIT Kavli Institue for Astrophysics and Space Research\\77 Massachusetts Avenue, Cambridge, MA, 02139}
\email{schech@mit.edu}

\and

\author{M.W.~Bautz}
\affil{MIT Kavli Institue for Astrophysics and Space Research\\77 Massachusetts Avenue, Cambridge, MA, 02139}
\email{mwb@space.mit.edu}

\begin{abstract}
Measuring dark matter substructure within galaxy cluster haloes is a fundamental probe of the $\Lambda$CDM model of structure formation.  Gravitational lensing is a technique for measuring the total mass distribution which is independent of the nature of the gravitating matter, making it a vital tool for studying these dark-matter dominated objects.  We present a new method for measuring weak gravitational lensing flexion fields, the gradients of the lensing shear field, to measure mass distributions on small angular scales.  While previously published methods for measuring flexion focus on measuring derived properties of the lensed images, such as shapelet coefficients or surface brightness moments, our method instead fits a mass-sheet transformation invariant Analytic Image Model (AIM) to each galaxy image.  This simple parametric model traces the distortion of lensed image isophotes and constrains the flexion fields.  We test the AIM method using simulated data images with realistic noise and a variety of unlensed image properties, and show that it successfully reproduces the input flexion fields.  We also apply the AIM method for flexion measurement to Hubble Space Telescope observations of Abell 1689, and detect mass structure in the cluster using flexion measured with the AIM method.  We also estimate the scatter in the measured flexion fields due to the unlensed shape of the background galaxies, and find values consistent with previous estimates.

\end{abstract}


\section{Introduction}
Gravitational lensing is a well developed tool for measuring the mass distributions of galaxy clusters.  Maps of the lensing convergence (the projected surface mass density between the observer and the source object scaled to the critical lensing density) in galaxy clusters, have been constructed using a variety of techniques for weak-field (or simply ``weak'') lensing observations \citep[e.g.,][]{kaiser+:95,hoekstra+:98,refregier+:03,kuijken:06}, multiple-image (or ``strong'') lensing observations \citep[e.g.,][]{kneib+:93,broadhurst+:05,diego+:05,coe+:10a}, and combining both strong and weak lensing observations \citep[e.g.,][]{bradac+:05}.  Joint weak+strong lensing approaches have the advantage of utilizing measurements of a broader range of lensing field strengths to produce a unified mass map.  Lensing is a vital tool for measuring the total masses and mass distributions of galaxy clusters, as it directly probes the dark matter which dominates the mass budget of these large cosmological objects.

While most work to date has focused on measuring image displacements and linear image distortions (strong lensing and weak lensing shear, respectively), gradients in the shear field which are significant on the scale of the lensed galaxy images produce second-order lensing distortions called ``flexion.''  These lensing effects, which can be decomposed into a comatic distortion (1-flexion) and a trefoil distortion (3-flexion), are probes of small scale structures in the mass distribution.

The two primary methods for measuring flexion in the current literature are shapelets \citep{goldberg+:05,massey+:07,leonard+:07} and surface brightness moments/Higher Order Lensing Image Characteristics \citep[HOLICs;][]{irwin+:06,okura+:07,schneider+:08}.  The shapelet method is, in a certain sense, a refinement of moment methods, since shapelet coefficients are measured by calculating a specific combination of surface brightness moments (corresponding to Hermite polynomials) using Gaussian weights.  The inversion of shapelet coefficients (or simply ``shapelets'') to estimate the lensing fields utilizes a Taylor series expansion of the surface brightness profile to linearize the lensing transformation with respect to the shapelets basis and takes advantage of the well-described mathematics of the quantum simple harmonic oscillator.  This linear approximation of the surface brightness profile is how the shapelets method fundamentally differs from moment methods, which use ratios of measured moments to approximate the flexion fields.  In both approaches, assumptions on the unlensed surface brightness profile are made in order to measure the lensing fields.  These assumptions are that odd-order shapelets/moments in observed images are sourced by lensing flexion alone, meaning that the intrinsic two-dimensional skewness of unlensed galaxies is assumed to be, on average, zero. The magnitude of the flexion scatter from intrinsic galaxy profile shapes is an important quantity to understand, as it sets the finite resolution of flexion mass measurements.  This limit is analogous to how the distribution of intrinsic ellipticities limits the spatial resolution of weak lensing shear mass measurements.

In each of the current flexion measurement methods, there are systematic issues that must be addressed.  Shapelet-based lensing measurements have been shown to be unreliable and biased for large shears \citep{melchior+:10}.  This is a significant concern for measuring flexion, since the shear can not be assumed to be small in the regime where flexion fields are measurable.  Moment-based measurements of flexion rely on high-order surface brightness moments, making the strength of the lensing signal extracted very sensitive to the window functions used in the moment calculations: though the flexion symmetries under coordinate transformations are matched by third-order moments of the surface brightness, the normalization of the moments required to isolate the flexion signal includes both fourth- and sixth-order moments \citep{okura+:08}.  Characterizing the noise properties of these moments and how the noise propagates into uncertainties in the measured flexion fields is a complex problem.  Additionally, the implementations of both these methods and their application to observational data have not addressed the well-established mass-sheet degeneracy \citep{gorenstein+:88,saha:00} in a consistent manner, as discussed by \cite{schneider+:08}.  These issues are not insurmountable, but further work needs to be done to refine flexion measurement, and alternative approaches for measuring flexion can provide a valuable insight to the strengths and weaknesses of each method.

In this paper we introduce a new method for weak lensing flexion measurement.  Instead of measuring derived quantities (such as weighted surface brightness moments, as in both the shapelets and HOLICs methods), we instead fit the lensed galaxy objects with a parameterized, Analytic Image Model (AIM) which is invariant to the mass-sheet degeneracy.  By comparing model images to the data image in ``pixel-space'' and optimizing a figure of merit over a reasonable range of model parameter values, we constrain the flexion fields.  This method has the advantage that surface brightness errors are well understood (and typically Gaussian), thus the optimization algorithm can provide reliable estimates of the errors on the best-fit parameters.  Uncertainties in the flexion measured by shapelet and moment methods are quantified on average, rather than for each individual object, and primarily estimate the uncertainty in the mass reconstruction instead of the uncertainty in the measured flexion.  Direct error estimates for each component of the 1-flexion and 3-flexion values from each individual lensed object is a very desirable property, as it allows us to accurately weight the flexion measurements from each object in mass reconstructions.

This paper is structured as follows: \S\ref{sec:flexionintro} reviews the basic flexion formalism used throughout the paper. \S\ref{sec:aimintro} describes the the principle of the AIM method and the specific implementation used here. \S\ref{sec:aimtest} describes the procedure used to test the AIM method on simulated data images, validating the accuracy of the fitting procedure and the accuracy of the error estimates.  \S\ref{sec:realdata} describes the Hubble Space Telescope Advanced Camera for Surveys Wide Field Camera (HST/ACS-WFC) observations of the galaxy cluster Abell 1689, which is used as a proving ground for the AIM method in a real data application.  We present detection of the substructure in A1689 using only 1-flexion measured with the AIM method, and the structure agrees with previously measured lensing maps.  In \S\ref{sec:intrinsic} we discuss our estimate of the scatter in flexion measurements due to the intrinsic (unlensed) shape of galaxies.

\section{Flexion Formalism}\label{sec:flexionintro}
The distortion of images by gravitational lensing is described by a coordinate transformation between positions $\theta$ in the image plane and positions $\beta$ in the unlensed source plane
\begin{equation}\label{eq:exactlensing}
	\beta=\theta-\nabla\psi(\theta),
\end{equation}
with
\begin{equation}
	\psi(\theta)=\frac{1}{\pi}\int\ d^2\theta'\ \kappa(\theta')\ln|\theta-\theta'|.
\end{equation}
We use complex notation here, where coordinates and lensing fields are denoted by complex quantities $x=x_1+ix_2$, and the derivative operator, $\nabla=\partial_1+i\partial_2$, is also complex.  $\kappa$ and $\psi$ are real-valued, but all other lensing quantities have non-trivial imaginary components.

\subsection{Second-Order Local Lensing}
In the case of weak lensing, the transformation in Equation \ref{eq:exactlensing} is expanded in a Taylor series valid in the neighborhood of the position of the image ($\theta_0$ in the image plane, $\beta_0$ in the source plane) in question.  To include the flexion fields, this expansion must extend to quadratic order, and yields
\begin{equation}\begin{aligned}\label{eq:lensingexpansion}
\beta_0+\beta=\	&\theta_0-\nabla\psi|_{\theta_0}+\\
				&(1-\kappa)\theta-\gamma\theta^*-\\
				&\frac{1}{4}\fflex^*\theta^2-\frac{1}{2}\fflex\theta\theta^*-\frac{1}{4}\gflex(\theta^*)^2.
\end{aligned}\end{equation}
The first-order lensing fields (convergence and shear) are $\kappa=\frac{1}{2}\nabla\nabla^*\psi$ and $\gamma=\frac{1}{2}\nabla^2\psi$; the-second order fields (flexion) are $\fflex=\nabla\kappa=\nabla^*\gamma$ and $\gflex=\nabla\gamma$.  Throughout this paper we will refer to $\fflex$ as 1-flexion and $\gflex$ as 3-flexion, refering to the spin symmetry of the fields with respect to coordinate rotations.  Elsewhere in literature, these lensing fields are called ``first flexion'' and ``second flexion'', respectively.  We prefer the spin-based notation because it indicates a physical property of the flexion fields, rather than an arbitrary ordering.  All derivatives of the lensing potential are evaluated at $\theta_0$.  This notation follows that of \cite{schneider+:08}, which simplifies the tensor notation of \cite{goldberg+:05}.

When considering a single lensed image, the constant terms in Equation \ref{eq:lensingexpansion} are degenerate and unmeasureable, and can be neglected in favor of small deviations $\theta$ about the observed image center $\theta_0$.  This gives a second-order local lensing equation
\begin{equation}\begin{aligned}\label{eq:flexionlensing}
\beta=	&(1-\kappa)\theta-\gamma\theta^*-\\
		&\frac{1}{4}\fflex^*\theta^2-\frac{1}{2}\fflex\theta\theta^*-\frac{1}{4}\gflex(\theta^*)^2.
\end{aligned}\end{equation}
This local lensing equation is sufficient for producing ``arced'' images from intrinsically circular or elliptical ones.  It is valid in the regime where the dimensionless products of the image size and the flexion fields, $a_I|\fflex|$ and $a_I|\gflex|$, are small with respect to unity.  The effect of flexion becomes significant when $a_I|\fflex|$ and/or $a_I|\gflex|$ are small but non-negligible corrections to the linearized lensing equation.  If the lensing flexion is significant, the weak lensing approximations, $\kappa\ll1$, $|\gamma|\ll1$, and $\beta(\theta)$ is linear, are typically no longer valid.  This has significant implications for characterizing the lensing transformation in regards to the mass-sheet degeneracy.

\subsection{Mass-Sheet Invariance}
The mass-sheet degeneracy is present in all local lensing measurements \citep{gorenstein+:88,saha:00}.  Any set of lensing fields $\{\kappa,\gamma,\fflex,\gflex\}$ estimated from the analysis of lensed images at a single redshift can only be constrained to a family of solutions $\{\kappa',\gamma',\fflex',\gflex'\}$, where $\kappa'=\lambda\kappa+(1-\lambda)$, $\gamma'=\lambda\gamma$, $\fflex'=\lambda\fflex$, and $\gflex'=\lambda\gflex$ are equally valid solutions for any non-trivial value of $\lambda$.  This degeneracy corresponds physically to an unconstrained source plane scale.  Multiply-imaged (strongly lensed) sources observed at multiple redshifts are required to constrain the source plane for those objects, fixing the constant terms in Equation \ref{eq:lensingexpansion} and breaking the mass-sheet degeneracy \citep{falco+:85,schneider+:95}.

There are combinations of the lensing fields which are invariant to mass-sheet transformations.  These are the reduced shear $g$, reduced 1-flexion $\Psi_1$, and reduced 3-flexion $\Psi_3$.  By referring to a rescaled, fiducial source plane where $\beta'=\beta/(1-\kappa)$, rather than the true source plane, the lensing transformation becomes
\begin{equation}\label{eq:reducedflexionlensing}
	\beta'=\theta-g\theta^*-\frac{1}{4}\Psi_1^*\theta^2-\frac{1}{2}\Psi_1\theta\theta^*-\frac{1}{4}\Psi_3(\theta^*)^2,
\end{equation}
where $g=\gamma/(1-\kappa)$, $\Psi_1=\fflex/(1-\kappa)$, and $\Psi_3=\gflex/(1-\kappa)$.  This notation is similar to that of \cite{schneider+:08}, though we do not absorb a factor of 1/4 into the definition of $\Psi_1$ and $\Psi_3$ to better match our parameter values to the flexion estimators of \cite{er+:10}.  This formulation quantifies the lensing transformation in terms of three complex variables rather than three complex and one real, casting the lensing transformation in terms of mass-sheet transformation invariant fields only.  Each of the three reduced lensing fields is specifically defined and measurable for single images, though determining $\kappa$ from these reduced lensing fields requires a constraint of the mass-sheet degeneracy.  From this point on we will use the transformation in Equation \ref{eq:reducedflexionlensing} and drop the prime notation from $\beta'$ simplicity.

\section{Measuring Flexion with AIM}\label{sec:aimintro}
In this section we describe our alternative to shapelets and moments for measuring weak lensing flexion.  Rather than measuring either set of derived quantities from the observed images and making assumptions on their properties in the (unobservable) unlensed image, we instead forward-model the observed image analytically.  By assuming a parameterized ansatz for the unlensed image and applying the quadratic lensing equation in Equation \ref{eq:reducedflexionlensing}, a model image is generated and compared to the observed, lensed image.  As we show, the ansatz need not be exactly matched to the true galaxy shape: an elliptical Gaussian is sufficient for most objects.  This forward-modeling approach is what sets the AIM method apart from shaplet and moment based methods, which instead attempt an inversion of image characteristics to measure 1- and 3-flexion.

\subsection{The Analytic Image Model Method}\label{sec:aimdef}
As a consequence of Liouville's theorem, the surface brightness of an astrophysical source is conserved by gravitational lensing.  This means that the observed surface brightness $I_{\text{obs}}$ at an image-plane position $\theta$ in terms of the source plane surface brightness $I_{\text{src}}$ is 
\begin{equation}
	I_{\text{obs}}\left[\theta\right]=I_{\text{src}}\left[\beta(\theta)\right],
\end{equation}
where $\beta(\theta)$ is the lensing coordinate transformation.  If we assume that the intrinsic, unlensed surface brightness profile can be well-described by a set of model parameters $\{p_{\text{int}}\}$, then the lensed model image is defined to be
\begin{equation}
	I_{\text{AIM}}\left[\theta;p_{\text{int}},p_{\text{lens}}\right]=I_{\text{src}}\left[\beta(\theta,p_{\text{lens}}),p_{\text{int}}\right],
\end{equation}
where $p_{\text{lens}}$ is a set of parameters characterizing the lensing transformation.  For measuring flexion with such an analytic image model, the parameter set $p_{\text{lens}}=\{g_1,g_2,\Psi_{11},\Psi_{12},\Psi_{31},\Psi_{32}\}$ defines this transformation.  Here $g_1$, $\Psi_{11}$, and $\Psi_{31}$ are the real parts, and $g_2$, $\Psi_{12}$, and $\Psi_{32}$ are the imaginary parts of $g$, $\Psi_{1}$, and $\Psi_{3}$, respectively.

The AIM is optimized by minimizing the figure of merit
\begin{equation}\label{eq:chisq}
	\chi^2(p_{\text{int}},p_{\text{lens}})=\sum_{n}\frac{\left(I_{\text{obs}}(\theta^{(n)}) - I_{\text{AIM}}(\theta^{(n)};p_{\text{int}},p_{\text{lens}})\right)^2}{\sigma_n^2},
\end{equation}
where $\theta^{(n)}$ is the image-plane position of the $n^{\text{th}}$ pixel and $\sigma_n^2$ is an estimate of the variance in that pixel's value.  The parameter set which minimizes this figure of merit is our estimate of the true set of intrinsic and lensing parameters.

\subsection{Implementation}
We implement the AIM method using a Levenberg-Marquardt minimization algorithm called \texttt{MPFIT} \citep{markwardt:09}. \texttt{MPFIT} is written in the Interactive Data Language\footnote{IDL is produced by ITT Visual Information Solutions: \href{http://www.ittvis.com}{http://www.ittvis.com}} (IDL).  Each model parameter is restricted to a fixed range of values encompassing the plausible span for that parameter.  The range for each parameter is described below and listed in Table \ref{tab:pars}.

The Levenberg-Marquardt minimization algorithm yields not only best-fit parameters, but also a full covariance matrix $(C_{jk})$ for all parameter pairs.  Both the variance estimate for each parameter ($\sigma^2_j=C_{jj}$) and the correlation matrix ($\rho_{jk}=C_{jk}/\sqrt{C_{jj}C_{kk}}$) provide additional indicators for the quality of the fit.  The variance estimates directly measure the constraint that the AIM minimization places on each model parameter for each object analyzed, and the correlation matrices can be used to evaluate and improve the model parametrization we use.

Models with many parameters (this implementation has twelve) can often have significant parameter degeneracies, making the correlation matrix obtained from the fitting a valuable tool for evaluating the specific model parametrization.  Large correlation coefficients ($|\rho_{jk}|\sim1$) indicate significant degeneracies which can prevent the minimization algorithm from efficiently and accurately converging.  

There is a lensing-parameter/shape-parameter degeneracy which is well known to the lensing community: the shear/ellipticity degeneracy utilized in standard weak lensing studies to measure shear.  Our approach to mitigating this degeneracy, fixing the shear parameters during fitting, will be discussed in \S\ref{sec:shearellip}, and the simulations described in \S\ref{sec:aimtest} confirm that there are no other strong degeneracies.  Modest correlations $|\rho_{jk}|\sim0.7$ exist for some lensed images and parameter combinations, but even in these instances convergence is robust and accurate.

The lensing transformation is characterized by six variables: the three complex, reduced lensing fields $g$, $\Psi_1$, and $\Psi_3$ from Equation \ref{eq:reducedflexionlensing}.  The range allowed for each these parameters is listed in Table \ref{tab:pars}.  We fix the shear during fitting to address the degeneracy between the shear and the intrinsic ellipticity, and so do not set explicit limits on the range of values for the two shear parameters.  \S\ref{sec:shearellip} discusses this point in more detail.

Though it is well-known that intrinsic galaxy profiles are not Gaussian \citep[see, e.g.,][and references therein, for a review of non-Gaussian galaxy shapes]{graham+:05}, we assume an elliptical Gaussian model for the unlensed images.  This is jusitified because the goal of the AIM method is not to match the profile of the galaxy image exactly.  Rather, the goal is to model the distortion by lensing of the galaxy isophotes.  A Gaussian profile effectively acts as a simple window function that identifies the isophotes with a minimal set of parameters.  The assumption that the unlensed isophotes are elliptical is a standard lensing assumption.  For an elliptical unlensed image, the isophotal ellipses will be similarly distorted for a variety of intrinsic surface brightness profile slopes.  Simulations which assumed \sersic\ profiles for the unlensed image model failed to converge at a much higher rate $\sim25\%$ and yielded flexion estimates with errors nearly an order of magnitude greater.  This is due to degeneracies between the normalization, size and \sersic\ index parameters.

With an elliptical Gaussian ansatz for the unlensed image, there are twelve model parameters: six for the unlensed profile and six for the lensing transformation.  The Gaussian parameters are combined into two real-valued variables, $\log S_0$ and $\alpha$; and two complex variables, $\theta_c$ and $\epsilon$.  In this parametrization, the unlensed surface brightness profile is defined by
\begin{equation}
	I_{\text{src}}\left[\beta\right]=\frac{S_0}{2\pi\alpha^2}\exp\left(-\frac{r(\beta)^2}{2\alpha^2}\right),
\end{equation}
with
\begin{equation}
	r(\beta)^2=(1+\epsilon\epsilon^*)\beta\beta^* - \beta^2\epsilon^* - (\beta^*)^2\epsilon.
\end{equation}
The image scale and complex ellipticity $\epsilon$ are defined in terms of the semi-axes and position angle ($A$, $B$, and $\phi$) as
\begin{equation}
	\alpha=\sqrt{AB};\quad\epsilon=\frac{A-B}{A+B}e^{2i\phi}\quad(A\geq B).
\end{equation}
The center position of the image is included by defining
\begin{equation}
	I_{\text{AIM}}\left[\theta\right]=I_{\text{src}}\left[\beta(\theta-\theta_c)\right].
\end{equation}
We define the center position in the image plane for a better \emph{a priori} limit on the allowed parameter range for the center position, and the origin of the source plane coordinates is defined to be at the center of the unlensed image.  Note that this center position does not refer to the image centroid, but rather the location in the image plane where the peak surface brightness of the source plane is mapped.  Table \ref{tab:pars} lists the parameter ranges allowed during minimization, which are chosen to be very broad and thus prevent parameters from reaching these limits.  However, if a parameter does happen to reach the limits while optimizing, that fit is flagged as a non-convergence.

In general, there will be a non-negligible background surface brightness and a finite point-spread function (PSF) size.  Since these observational effects are not well constrained by the individual lensed galaxy images, we take them as given for the AIM optimization.  In practice, they will be determined externally to the AIM fitting; see \S\ref{sec:realdata} for our approach.

\subsection{Shear/Ellipticity Degeneracy}\label{sec:shearellip}
As is well known from weak lensing studies, there is a complete degeneracy between shear and ellipticity.  Assuming only a linear lensing equation (neglecting flexion), the observed ellipticity is given by \citep{bartelmann+:01}:
\begin{equation}\begin{aligned}\label{eq:shearellip}
	\epsilon_{\text{obs}}	&=\frac{\epsilon+g}{1+\epsilon g^*}\quad\text{for }|g|<1,\\
							&\\
							&=\frac{1+\epsilon^* g}{\epsilon^*+g^*}\quad\text{for }|g|>1.\\
\end{aligned}\end{equation}
Except in the limit of extreme image distortions, this degeneracy prevents the shear from being well-constrained by a single lensed image.  This is why weak lensing shear studies must average the ellipticities of nearby galaxies in order to infer the lensing shear.  For the AIM measurements of flexion presented here, the presence of this degeneracy means that the four-dimensional parameter volume spanned by $\epsilon$ and $g$ only has two constraints ($\epsilon_{\text{obs}}$).  In order to accurately constrain flexion using the $\chi^2$ minimization, we must reduce this part of the AIM parameter space by two dimensions to prevent the minimization algorithm from converging to degenerate local minima rather than finding the true global minimum.  We do not expect that the specific constraint on $\epsilon$ and $g$ that we choose will affect the flexion estimation, since the flexion fields distort images with different symmetries than shear and ellipticity (spin-1 and spin-3 flexion, versus spin-2 shear and ellipticity); and indeed, the moderate correlation coefficients between shear/ellipticity parameters and flexion parameters in our simulations (described in \S\ref{sec:aimtest}) bear out this expectation.

To remove this degeneracy we choose to fix the two shear parameters to an assumed input model.  This eliminates two dimensions from the parameter space, drastically reduces the convergence time of the fitting algorithm (by a factor of $\sim10$), and allows the flexion to be accurately and precisely constrained.  In simulations where both the ellipticity and the shear are allowed to vary freely, the flexion parameter values returned by the minimization place no significant constraint on the lensing flexion, and nearly all the parameter values are consistent with zero flexion within the error estimates derived from the parameter covariance matrix.

We choose to fix the shear, rather than fixing the ellipticity, because it is possible to select a specific \emph{a priori} shear model which is observationally motivated by additional data on the galaxy cluster in question.  The shear can be estimated from mass proxies (e.g.~standard weak lensing shear analysis, X-ray temperature or luminosity, or optical richness) and the assumption of a standard mass profile (e.g., a non-singular isothermal sphere or NFW profile).  Significant departures from the input shear model would be measurable as departures from the expected distribution of ellipticity magnitudes and orientations in the lensed galaxies.  

Flexion measurements are a natural addition to joint weak+strong lensing mass mapping methods, as flexion directly constrains the mass gradients.  Mass reconstruction formalisms, such as that of \cite{bradac+:05} or the non-grid-based method of \cite{deb+:08}, can be easily modified to utilize data which further constrains the lensing potential.  As a less-sophisticated method which does not use direct shear estimates, an iterative process of integrating flexion measurements could be used to correct the assumed input shear model (and thus the mass model) for the effects of lens substructure.  The resulting updated shear model could then be used as an input to the next iteration of flexion fitting, and the process repeated until the shear model converges.  This type of procedure would be a test of the self-consistency of the measured shear and flexion fields; however it is beyond the scope of this work.  The simulations we present in \S\ref{sec:aimtest} confirm that our basic approach to the shear/ellipticity degeneracy (fixing the shear and fitting for the ellipticity) is adequate for successfully measuring the lensing flexion, even when significant errors are present in the assumed shear field.

\section{Testing the AIM Method}\label{sec:aimtest}
We test the efficacy of the AIM method for measuring flexion using simulated data images.  Image noise and pixel scale are chosen to match the Abell 1689 data described in \S\ref{sec:realdata}.  Data images are generated by randomly selecting intrinsic shape and lensing parameter values, assuming a specific intrinsic galaxy profile and applying the quadratic lensing transformation.  These images are circularly windowed such that the observed image centroid is situated in the center of the simulated data image, to emulate images that would be extracted from real data.  The AIM method is used to determine best-fit parameter values.  The lensing parameters are either selected uniformly from fixed ranges or from a Singular Isothermal Sphere (SIS) lens model constrained to the same volume of lensing parameter space.  A set of 1000 simulated images are generated to explore the AIM parameter space for each lensing parameter selection method.

\subsection{Intrinsic Shape Parameter Selection}
Each intrinsic image follows a \sersic\ profile,
\begin{equation}
	I\propto\exp(-0.5(r/\alpha)^{1/n}), 
\end{equation}
with the index $n$ uniformly selected from the range 0.2 to 5.  For reference, $n=0.5$ is a Gaussian, $n=1$ is exponential, and $n=4$ is a de Vaucouleurs' profile.  The intrinsic size parameter $\alpha$ is drawn from a uniform range between $0\farcs25$ and $1\farcs25$.  The normalization $\log S_0$ (in arbitrary units) is selected uniformly from a range such that the object would be a 2-$\sigma$ detection or better.

The ellipticity magnitude is drawn from a Gaussian distribution of the form
\begin{equation}
	p(\epsilon)=\frac{1}{\pi\sigma^2\left(1-\exp(-1/\sigma^2)\right)}\exp\left(\frac{-|\epsilon|^2}{\sigma^2}\right),
\end{equation}
with $\sigma=0.2$, but truncating the distribution at $|\epsilon|=0.9$, as in \cite{schneider:96}.  The phase of the ellipticity is uniformly selected to be from $0$ to $2\pi$, randomly orienting the unlensed ellipse.  The center position $\theta_c$ is selected such that the image centroid is located at the center of the data image.

The AIM method assumes that the intrinsic profile is Gaussian, while the simulated data images explicitly include non-Gaussian \sersic\ profiles.  As a result, the best-fit values for $\log S_0$ and $\alpha$ will not match well to the input values if the \sersic\ index differs significantly from 0.5.  However, the shape of the isophotes is still constrained and the other intrinsic shape parameters (ellipticity and center position) can be well fit.  The exact true values of $\log S_0$ and $\alpha$ for the unlensed image need not be determined for an accurate lensing analysis.

\subsection{Lensing Parameter Selection}
Lensing parameters are randomly selected using one of two methods.  The first method is to simply draw each parameter from a uniform distribution in a volume of parameter space.  For this uniform selection, we restrict the parameters to $|g_n|<0.5$, and $|\Psi_{mn}|<0.1$ arcsec$^{-1}$.  Images are generated using the quadratic lens equation.

The second selection method involves using the uniformly selected parameters to choose a specific SIS lens model.  We do this to test the efficacy of the AIM method on images which have been lensed analytically from a well-defined lensing potential, rather than restricting the lensing to be strictly quadratic.  For an image located at a radius $r$ and position angle $\phi$ from the center of a SIS, the lensing fields are
\begin{eqnarray}
	g=-\frac{\theta_E}{2r-\theta_E}\exp(2i\phi), \\
	\Psi_1=-\frac{\theta_E}{r(2r-\theta_E)}\exp(i\phi),
\end{eqnarray}
and
\begin{equation}
	\Psi_3=\frac{3\theta_E}{r(2r-\theta_E)}\exp(3i\phi), 
\end{equation}
where $\theta_E$ is the Einstein radius of the SIS.  From the uniformly selected lensing parameters, we calculate $r/\theta_E$ from $|g|$; $\phi$ from $\Psi_{12}/\Psi_{11}$; and $r$ and $\theta_E$ are separately determined by combining $|\Psi_3|$ with $|g|$.  This allows us to select a uniform distribution of lensing parameters while requiring that the lensing field combinations be those of a SIS.

\subsection{Image Generation}
Simulated images are created and windowed to include only those pixels at a distance from the observed image centroid of a factor of 1.5 times the observed semi-major axis length $A$.  This window size is empirically determined to be sufficiently large to measure flexion accurately while being independent of ellipticity and not being so large as to include excessive sky noise.  This choice of $1.5A$ is motivated by similar concerns presented by \cite{goldberg+:07}.  For uniformly selected lens parameter images the simulated data images are generated using the AIM described in \S\ref{sec:aimdef}.  For SIS lens parameter images, the lensing transformation is known exactly, and so the images are analytically lensed without approximating the transformation to second order.

Each data image is convolved with a circular, Gaussian PSF with a size similar to that of the HST/ACS-WFC.  Gaussian-distributed pixel noise is included with a standard deviation chosen to match the dataset described in \S\ref{sec:realdata}.  We do not use a more accurate HST PSF model \citep[e.g.~TinyTim;][]{krist:93} because the effects of the PSF on the measured flexion are expected to be negligible due of the small PSF size (0\farcs08 FWHM) relative to the size of galaxies which we select for analysis ($>0\farcs24$ FWHM).  Assuming that the PSF has a shape corresponding to a 1-flexion $\fflex_{\text{PSF}}$, the effect on the measured 1-flexion scales as
\begin{equation}
	\fflex_{\text{induced}}\sim\fflex_{PSF}\frac{\alpha_{\text{PSF}}^4}{\alpha_{\text{PSF}}^4+\alpha_{\text{image}}^4}
\end{equation}
sharply damping out PSF-induced flexion for galaxies only modestly larger than the PSF \citep{leonard+:07}.  PSF ellipticity couples to the ellipticity and shear parameters, not to the flexion parameters.  The approximation of a Gaussian PSF reduces the computational intensity of the AIM method and is a valid simplification, though a more complex PSF model would need to be used for images where the PSF size were larger, such as for ground-based imaging.

\subsection{Initial Parameter Selection}
Initial parameter values for intrinsic shape parameters are determined from surface brightness moments of the data images.  The center position is initially set at the observed centroid of the simulated data image.  The flexion parameters are all initialized at zero, meaning that the ``first guess'' model image is always an elliptical image, and non-zero flexion measurements will be positive detections by the fitting algorithm.

As noted in \S\ref{sec:shearellip}, there is a degeneracy between the ellipticity and the shear, requiring that the shear parameters be fixed to some set value for the minimization to converge.  For these simulations, each shear parameter is fixed to a value deviated about the true value by a Gaussian distribution with a standard deviation of 0.2 (no units).  Accordingly, the initial ellipticity parameter values are adjusted using Equation \ref{eq:shearellip} so that $\epsilon_{\text{obs}}$ matches the ellipticity determined from the simulated data image moments.  This ensures that the observed ellipticity of the initial model image matches that of the data image.

\subsection{Simulation Results}
The AIM method converges to a minimum rapidly and consistently ($>99\%$ convergence, typically in less than 100 iterations).  The final figure of merit $\chi^2$ per Degrees of Freedom (D.o.F.) for converged fits have mean values of $\sim1.0$--$1.2$.  A typical data image has 500--1000 D.o.F.  Successful, converged fits are identified from the larger ensemble as those having  $\chi^2/\text{D.o.F.}<1.5$ and $\sigma(\Psi_{mn})>0.001$ arcsec$^{-1}$.  The latter condition is slightly couterintuitive, but it is useful for excluding fits which tightly constrain one or more flexion parameters to erroneous values in local minima of the $\chi^2$ surface.  We are able to identify this threshold because we know the true input flexion values.  Fits with one or more flexion parameter constrained this tightly by the minimization do not return accurate flexion estimates.

\begin{figure}[h]
\plottwo{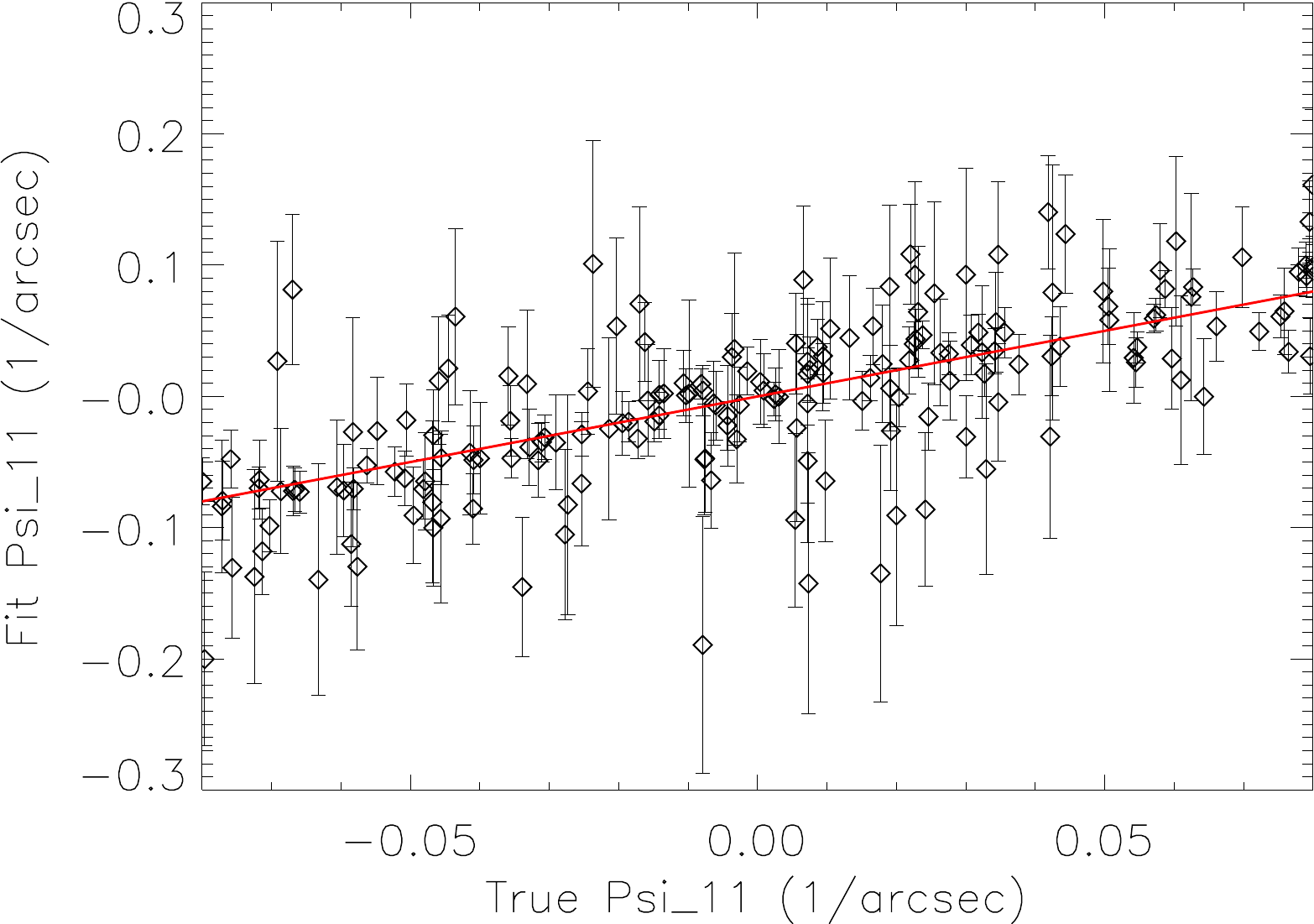}{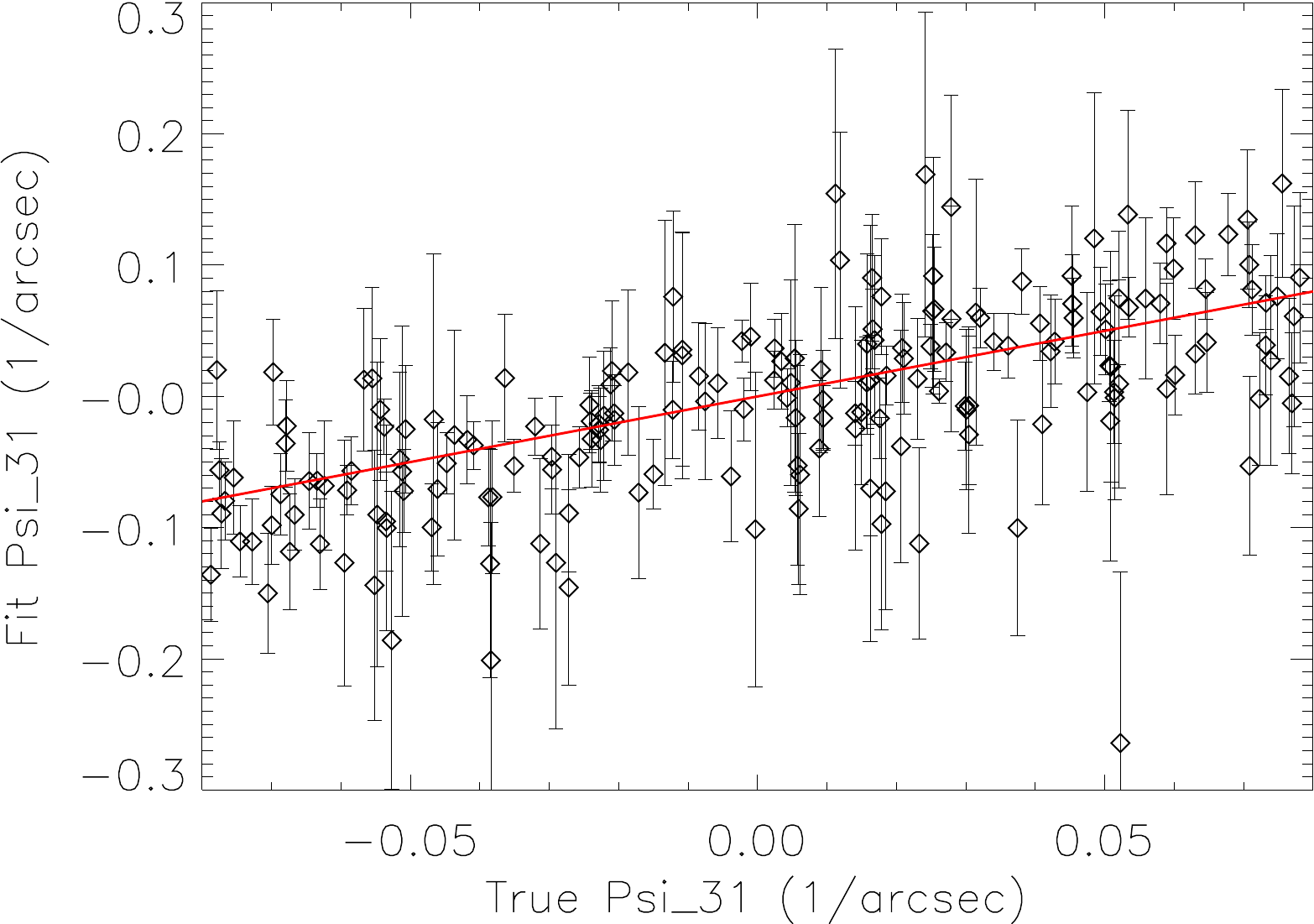}
	\caption{A comparison of fit and input flexion values from simulations for 1-flexion and 3-flexion.  We plot only one component of each flexion field, but the other is very similar.  The points in both panels are drawn from the same representative subset of the simulated objects, and the overlaid red line is a unity line, not a fit.  As is typical, the 3-flexion errors are larger and the fit values are more scattered.  The angular flexion units assume the HST ACS pixel scale.\label{fig:flexplot}}
\end{figure}

The error between the true and fit normalization and size parameters ($\log S_0$ and $\alpha$) are small, with RMS values of $0.1$ for $\log S_0$ and $0\farcs04$ for $\alpha$.  These parameters are positively correlated, with typical correlation coefficients of $\rho_{\log S_0/\alpha}\sim0.7$.  For low surface brightness \sersic\ input profiles with indices significantly different from the Gaussian value of 0.5, the errors in $\log S_0$ and $\alpha$ become exaggerated.  This scatter from model mismatching is expected, and residual images show that the central region of the image is not well estimated.  The flexion can still be accurately determined, as flexion primarily distorts the isophotes away from the image center.  This property is indicated by the absence of any significant correlations between the flexion parameters and either the normalization or size parameters.  The RMS center position error is $\sim0\farcs05$ (approximately one pixel).  The ellipticity is measured with an RMS error of $\sim0.2$, matching the distribution of the input shear values.  Similar simulations fixing the shear parameters to their true values yield an RMS error on the ellipticity of $\sim0.01$.  Figure \ref{fig:flexplot} shows a comparison of fit and input 1- and 3-flexion values from simulations.  As has been observed in earlier work \citep[e.g.][]{goldberg+:07}, there is more noise and scatter in measurements of 3-flexion.

Flexion parameters are also well-recovered by the AIM method.  Though there is some correlation between the flexion and ellipticity parameters, it is modest ($|\rho_{jk}|\lesssim0.7$) and does not preclude accurate recovery of the input flexion fields.  Typical error estimates from the flexion variances yield $\sigma(\Psi_{mn})\sim0.01$ arcsec$^{-1}$, and the RMS deviation from the input flexion values is $0.007$ arcsec$^{-1}$.  This indicates that for those fits selected from within the acceptable ranges of $\chi^2/\text{D.o.F.}$ and $\sigma(\Psi_{mn})$, the parameter variances provide an accurate estimate of the deviations of the measured flexion values from the true lensing fields.

\section{Application to A1689}\label{sec:realdata}
Flexion has been used to measure mass structure in two contexts to date: galaxy-galaxy flexion in field galaxies \citep{goldberg+:05,velander+:11} and in the galaxy cluster Abell 1689.  This massive cluster is one of the best-studied gravitational lenses, making it an ideal testbed for a new lensing analysis technique.  Additionally, it is the only galaxy cluster to date where flexion has been used to measure the mass distribution.  \cite{leonard+:07} used shapelets to measure flexion in A1689 and constrained mass structures with a parametric model.  \cite{okura+:08} used HOLICs to measure flexion and found similar structure using weak lensing shear and flexion measurements and a Fourier mass reconstruction.  In this section we apply the AIM method of flexion measurement to substructure detection in A1689 as a real-world validation of our method.

It is important to note that in both previous flexion analyses of A1689, the authors mistreat the mass-sheet degeneracy in some way.  As shown by \cite{schneider+:08}, this is not a valid approximation for the interior region of the cluster where $\kappa\not\ll1$.  This means that direct, quantitative comparisons between this paper and previous work are not feasible.  Instead, we look for qualitative morphological similarities between the structures detected, with the intent of performing a direct comparison of flexion measurement methods in future work.

\subsection{The Data}
The data were obtained from the Hubble Legacy Archive\footnote{The Hubble Legacy Archive is located at: \href{http://hla.stsci.edu/}{http://hla.stsci.edu/}} (HLA) and reduced using the standard HLA pipeline.  These data were originally obtained for proposal HST-GTO/ACS9289 (PI Ford).  The data include deep observations in F475W (9.6 ks), F625W (9.6 ks), F775W (11.8 ks), and F850LP (16.6 ks).  The observed field is $3\farcm4\times3\farcm4$, and 1\arcsec\ on the sky corresponds to 3.1 kpc at the redshift $z=0.187$ of the cluster, assuming a standard cosmology.

\subsection{Object Selection}
Candidate objects for flexion analysis are selected using SExtractor \citep{bertin+:96} in a two-pass strategy similar to those of \cite{rix+:04} and \cite{leonard+:07}, with some modifications.  The first pass is intended to select the large, bright, cluster member galaxies and foreground objects.  These objects are then removed from the image in a process we describe below.  The resulting ``cleaned'' image is then used in the second pass to identify the smaller, fainter, lensed background galaxy images.  In all cases, a co-added, four-filter image is used for object detection to increase the signal-to-noise ratio of faint object detections, and the individual filter images are used for photometry.  SExtractor uses the input data images and pixel weights to produce a background image, noise image, and object image for each of the four filters.  The background and noise images contain position-variable estimates of the background surface brightness and the total surface brightness variance across the field.  The object image is the same as the input image except that any pixels not determined to be associated with a detected object are set to zero.

In the first pass, we select objects of large area (area $>50$ pixel$^2$) to a low threshold (SNR $>1$).  This identifies object-associated pixels down to the background level, including the faint Intra-Cluster Light (ICL) associated with the central cluster galaxies.  The detected objects are matched with known foreground and cluster member objects from literature \citep[and the NASA/IPAC Extragalactic Database \footnote{NASA/IPAC Extragalactic Database (NED): http://nedwww.ipac.caltech.edu/}]{duc+:02,coe+:10b}.  All known objects at a spectroscopic redshift $z<0.2$ are selected to be cleaned from the image.  All pixels associated with any known object location, as determined in the SExtractor object image using a friends-of-friends algorithm, are replaced with Gaussian-distributed noise.  The object pixels are cleaned in the four-filter detection image and in each individual filter image as well.  The standard deviation of the noise, as determined by SExtractor and output to the noise image, is enhanced in the object pixels by a factor of 2.  This is to account for statistical error in the subtraction of the image pixels as well as any systematic error in the subtraction process.

The cleaned images are then re-run through SExtractor, though now allowing for slightly smaller objects (area $>25$ pixel$^2$) with a more exclusive noise threshold (SNR $>3$).  Note that a new noise image for use in the flexion analysis is created in the second pass.  This returns a catalog of objects for flexion analysis, along with initial parameter values for $\log S_0$, $\alpha$, and $\epsilon$.  Those objects with $\alpha<2$ pixels are removed from the catalog as they are either spurious detections of noise peaks, stars, or are PSF-dominated galaxies which will have unreliable flexion estimates.  Our flexion analysis catalog includes 764 objects in the field (a density of $\sim66$ arcmin$^{-2}$).

There is an additional surface brightness cut which is applied after the fitting, removing objects with a peak surface brightness of $S_0/(2\pi\alpha^2)<5\ e^-/s/$square-pixel.  This removes any remaining spurious noise detections and objects with low signal more effectively than using the SExtractor determined initial values.  Application of this cut before the flexion analysis step removes $\sim30\%$ of the usable, flexion-yielding objects in addition to the unsuitable objects.  This is because for detections which are noise peaks or excessively faint, the AIM fitting algorithm naturally reduces $\log S_0$ and increases $\alpha$ so that the surface brightness falls below this threshold.  For usable objects, there is a floor for $\log S_0$ and a ceiling for $\alpha$ beyond which the $\chi^2$ values become unfavorable, and so these faint objects remain in our sample and can be used to constrain the flexion fields.

\subsection{Flexion Fitting}
``Postage-stamp'' images of each of the objects selected for flexion analysis are excised from the F775W data image and windowed to within a circular radius of 1.5 times the SExtractor-determined semi-major axis size about the observed centroid position.  As noted in \S\ref{sec:shearellip}, the input shear parameters must be fixed for the fit to converge accurately.  We choose a non-singular isothermal sphere (NIS) as a model for the cluster shear.  The (non-reduced) shear and convergence for a NIS are
\begin{equation}
	\kappa(\theta)=\frac{\theta_E}{\sqrt{\theta\theta^*+\theta_C^2}}
\end{equation}
and
\begin{equation}
	\gamma(\theta)=\kappa(\theta)\theta^2\left[\frac{2\theta_c\sqrt{\theta\theta^*+\theta_c^2}-\theta\theta^*-2\theta_c^2}{(\theta\theta^*)^2}\right].
\end{equation}
We set $\theta_E=$49\farcs5 and $\theta_C=$17\arcsec, as measured by \cite{broadhurst+:05}.  As noted by Broadhurst et al., it is unclear whether a NIS model, a NFW model, or some other mass model best describes the large scale cluster potential, particularly given the amount of known substructure in the cluster, but the NIS model is consistent with previous data and is a simple model to use for the shear.  Since our approach focuses on measuring flexion rather than accurately modeling the shear, even with errors the assumed shear model is an adequate approximation, though certainly not a perfect representation of the true shear field.  As noted in \S\ref{sec:shearellip}, the ellipticity will compensate for errors in the shear model.  An aphysical model with the shear assumed to be uniformly zero across the entire field was also tested, with very few significantly different flexion measurements, though the median $\chi^2/$D.o.F.~was slightly elevated.  This supports our assertion that the flexion parameters are minimally correlated with either the shear or ellipticity parameters.

\begin{figure}[h]
	\plotone{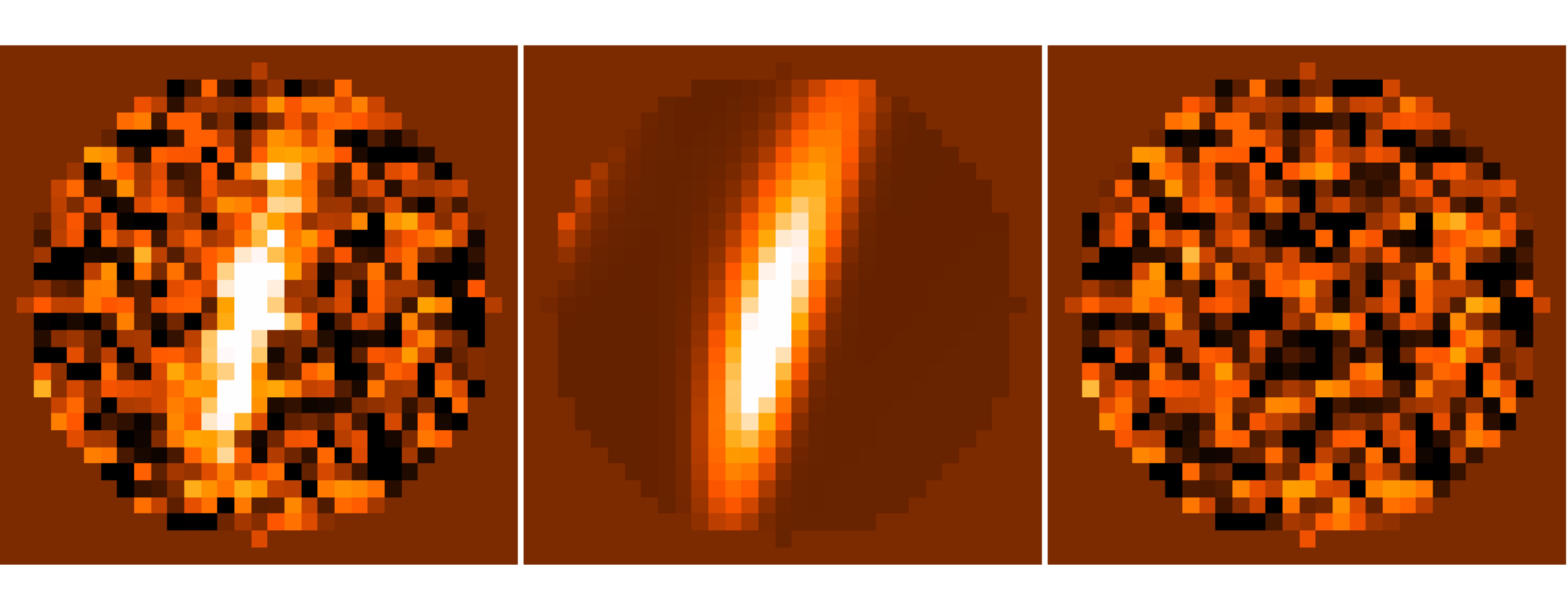}
	\plotone{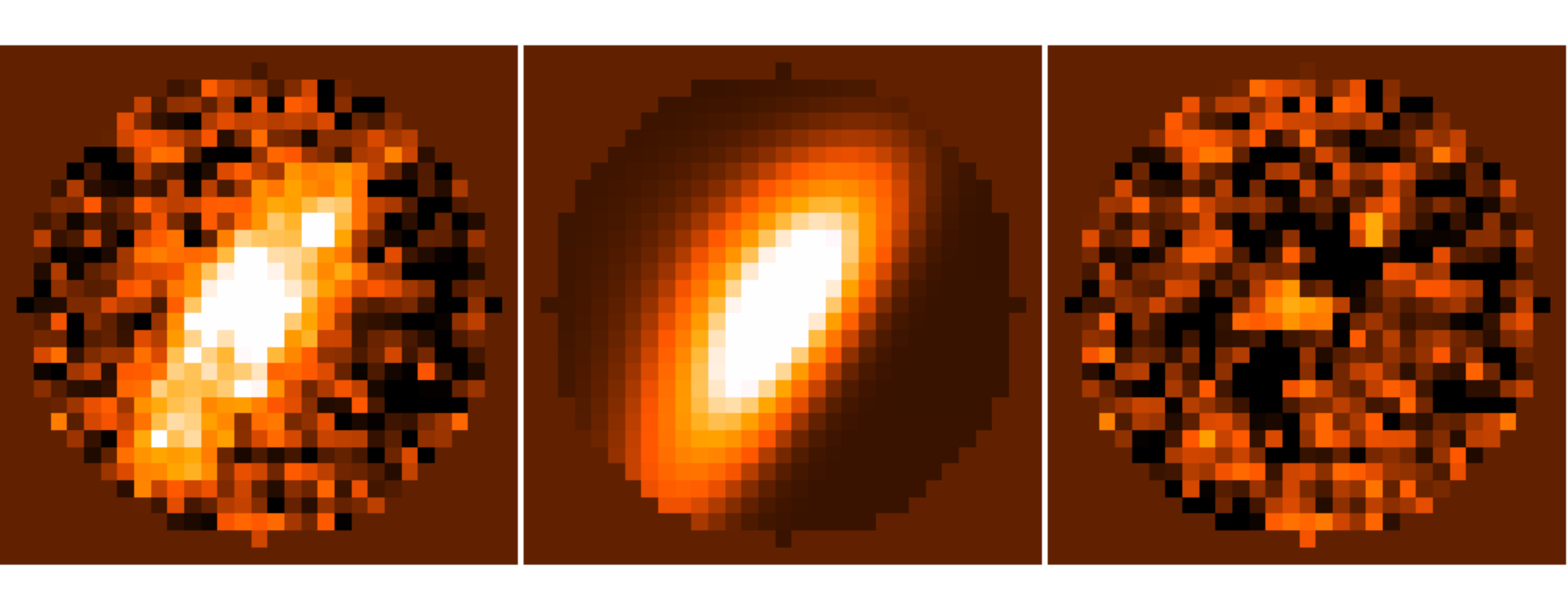}
	\plotone{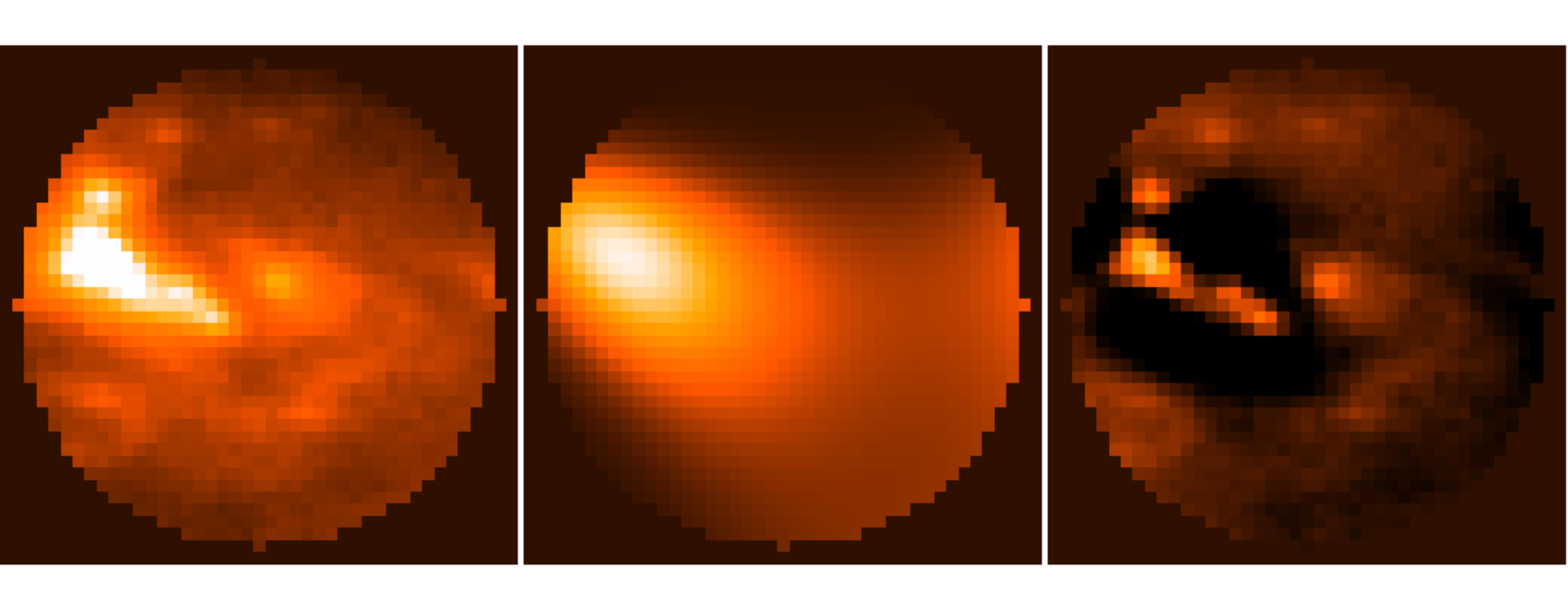}
	\caption{Example A1689 fit objects.  In all three rows, the left panel is the A1689 F775W data image used for the image analysis, the middle panel is the best-fit model image, and the right panel is the residual image.  The top row is a well-fit object with $\chi^2/$D.o.F.$<1.5$, the middle row is an object just below the threshold $\chi^2/$D.o.F.=1.5, and the bottom row is an object with $\chi^2/$D.o.F.$>1.5$, and thus rejected from the mass reconstruction.\label{fig:goodbadugly}}
\end{figure}

As in \S\ref{sec:aimtest}, we remove all fits with $\chi^2/$D.o.F.$>1.5$, and those with $\sigma(\Psi_{mn})<0.001$ arcsec$^{-1}$.  Figure \ref{fig:goodbadugly} shows three example fits: one well below the $\chi^2/$D.o.F. threshold, one just accepted, and one which was rejected.  The typical rejected objects are either faint objects located near a significantly brighter one, or extremely low surface brightness objects.  A final flexion catalog of 301 objects remained after these quality cuts were imposed.  We use these 301 objects to detect the mass structure in A1689.  For reference, Table \ref{tab:fitresults} lists the position and fit parameters for the 50 objects with the largest 1-flexion signal-to-noise ratio.

\subsection{Mass Structure Reconstruction}
We reconstruct the mass structure from the measured 1-flexion field, using a modified version of the mass reconstruction technique introduced by \cite{leonard+:10}.  Their method, which is similar to the aperture mass measurement techniques used for shear \citep[e.g.][]{schneider:96}, relates the weighted integral of the convergence on a circular aperture of radius $R$ about a center position $\theta_0$
\begin{equation}
	M_{ap}(\theta_0)=\int_{|\theta|\leq R}d^2\theta\ \kappa(\theta+\theta_0) w(|\theta|),
\end{equation}
to an integral of the 1-flexion:
\begin{equation}
	M_{ap}(\theta_0)=\int_{|\theta|\leq R}d^2\theta\ \fflex_E(\theta;\theta_0) Q_{\fflex}(|\theta|).
\end{equation}
The E-mode 1-flexion,
\begin{equation}
	\fflex_E(\theta;\theta_0)=\fflex(\theta)\frac{\theta^*-\theta^*_0}{|\theta-\theta_0|}, 
\end{equation}
is the component of the 1-flexion which is radially oriented from $\theta$ towards $\theta_0$.  The 1-flexion kernel $Q_{\fflex}$ is related to the aperture weight function $w$ by the relations
\begin{equation}
	Q_{\fflex}(x)=-\frac{1}{x}\int_0^x\ w(x')\ x'dx',
\end{equation}
and
\begin{equation}
	w(x)=-\frac{1}{x}Q_{\fflex}(x)-\frac{dQ_{\fflex}}{dx}.
\end{equation}
The kernel $Q_{\fflex}$ has units of angle, $\fflex$ has units of inverse angle, while $M_{ap}$ and $\kappa$ are unitless.  The weight function is constrained to go to zero smoothly at the aperture boundary and also to have the property that
\begin{equation}
	\int_0^R \ w(x')\ x'dx'=0.
\end{equation}
This latter restriction fixes the constant of integration to zero, eliminating the degeneracy $\kappa\rightarrow\kappa+\kappa_0$ from the mass reconstructions, where $\kappa_0$ is a constant offset.  Note that this is \emph{not} the mass-sheet degeneracy.  \cite{leonard+:10} assume that the flexion fields measured are the non-reduced flexion fields $\fflex$ and $\gflex$.

However, as noted above, we can not measure $\fflex$, and instead measure $\Psi_1$.  The relations between $M_{ap}$, $\kappa$, and $\fflex$, as well as those between $w$ and $Q_{\fflex}$ derive from the gradient relation between $\kappa$ and $\fflex=\nabla\kappa$.  $\Psi_1$ has its own gradient relation: $\Psi_1=\nabla K$, where
\begin{equation}
	K=-\ln|1-\kappa|.
\end{equation}

We define an aperture statistic similar to $M_{ap}$ using the E-mode reduced 1-flexion, $\Psi_{1E}$, defined analogously to $\fflex_E$.  This statistic is
\begin{equation}\label{eq:apmass}
\begin{aligned}
	K_{ap}(\theta_0)	&=\int_{|\theta|\leq R}d^2\theta\ K(\theta+\theta_0) w(|\theta|)\\
						&=\int_{|\theta|\leq R}d^2\theta\ \Psi_{1E}(\theta;\theta_0) Q_{\Psi_1}(|\theta|),
\end{aligned}
\end{equation}
where $Q_{\Psi_1}$ is related to $w$ in the same way that $Q_{\fflex}$ is.  Neglecting the difference between $\fflex$ and $\Psi_1$ (and thus the difference between $M_{ap}$ and $K_{ap}$) is a likely contributor to the anomalous B-mode convergence measured by \cite{leonard+:10}.  We use $K_{ap}$, calculated from our measured mass-sheet invariant 1-flexion, to detect mass structure without making any assumptions about the redshift distribution of the source objects.

We reconstruct $K_{ap}(\theta)$ from our flexion measurements using the following method.  The integral in Equation \ref{eq:apmass} can be converted to a discrete sum over $N$ flexion measurements simply:
\begin{equation}
	K_{ap}(\theta_0)=\sum_{n}\Psi_{1E}(\theta^{(n)};\theta_0)Q_{\Psi_1}(|\theta^{(n)}|)q_n.
\end{equation}
Here $q_n$ is a normalized weight factor given by the inverse square of the 1-flexion errors returned by the AIM fitting.

For the aperture weight function, we use the polynomial weights described in \cite{schneider+:98} and \cite{leonard+:10}.  These form a non-optimized, general family of polynomial functions characterized by a polynomial index $l$ and a maximum aperture radius $R$.  They are defined by the 1-flexion kernel
\begin{equation}
	Q(r)=-A_l\frac{2+l}{2\pi}\ r\left(1-\frac{r^2}{R^2}\right)^{1+l}
\end{equation}
and the aperture weight function
\begin{equation}
	w(r)=A_l\frac{(2+l)^2}{\pi}\left(1-\frac{r^2}{R^2}\right)^l\left(\frac{1}{2+l}-\frac{r^2}{R^2}\right),
\end{equation}
with the normalization factor
\begin{equation}
	A_l=\frac{4}{\sqrt{\pi}}\frac{\Gamma(7/2+l)}{\Gamma(3+l)}.
\end{equation}
chosen so that 
\begin{equation}
	\frac{1}{R^3}\int_0^RQ(r)2\pi rdr=1.
\end{equation}
Because the aperture radius $R$ can be freely selected, and the width of the kernel varied by changing $l$, this family of functions is sensitive to a wide range of mass structure scales.  Features which can be identified in several mass reconstructions with different aperture radii and different slopes can be associated with real structure and not noise peaks.

\begin{figure}[h]
	\plottwo{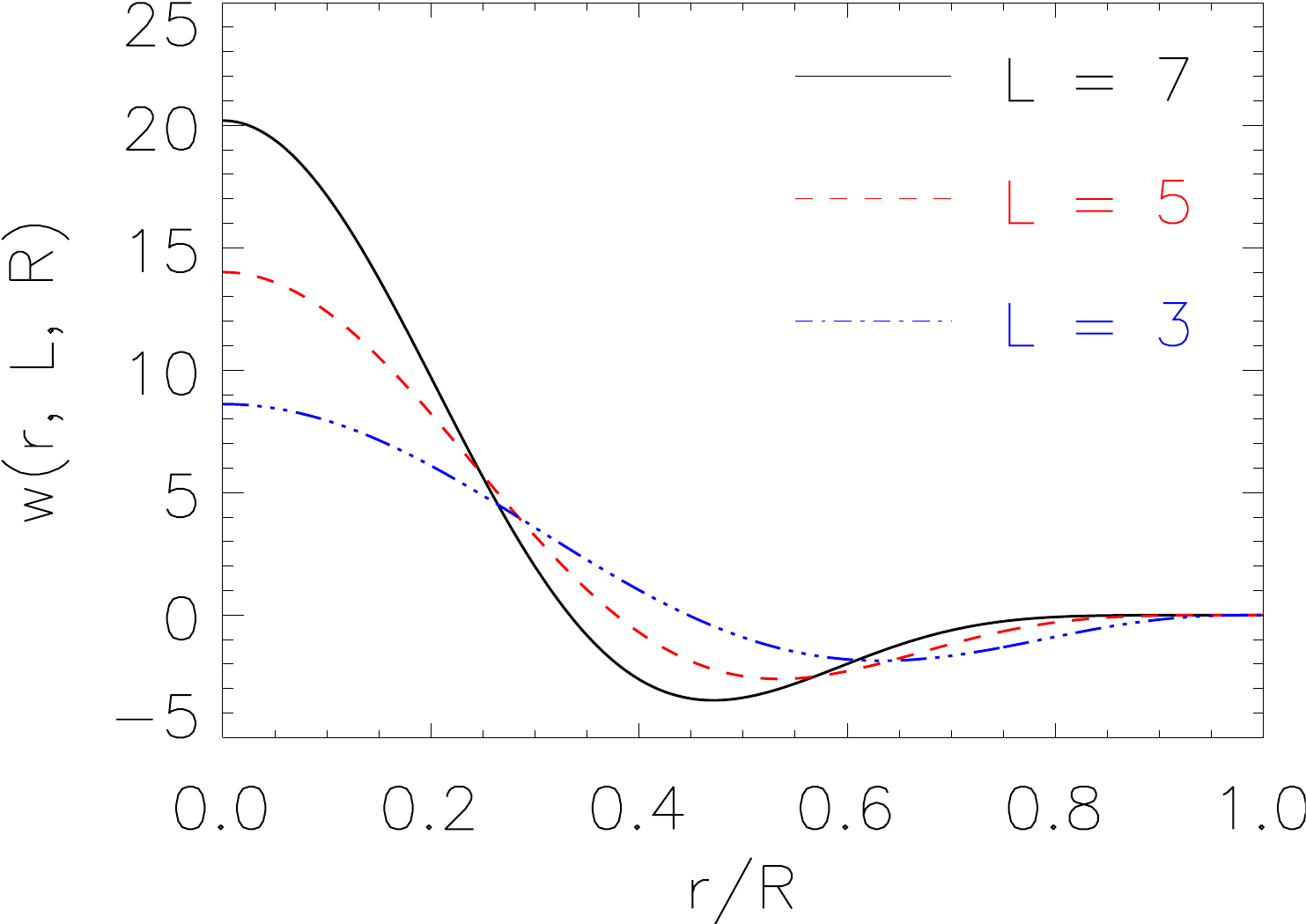}{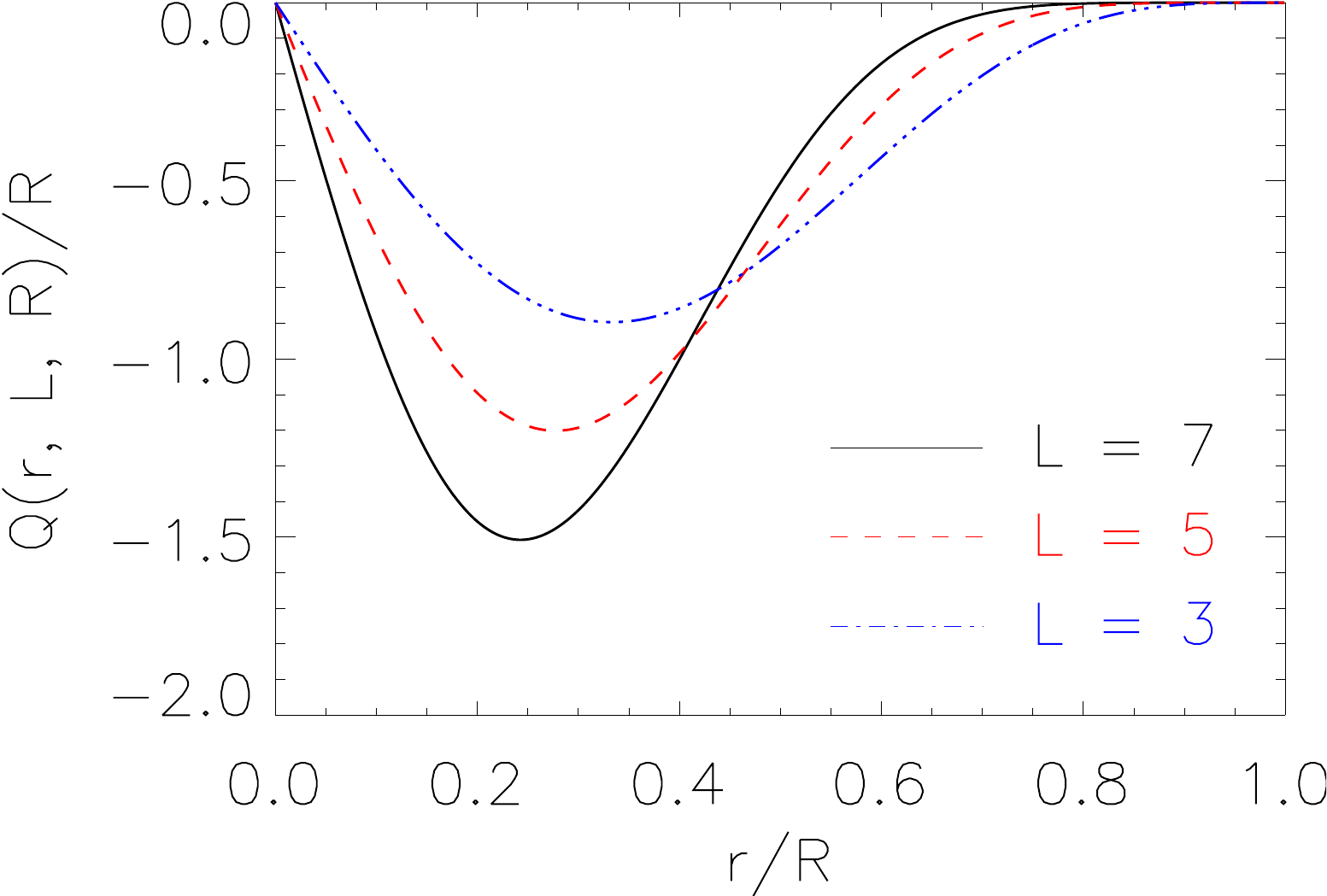}
	\caption{Aperture weight (left) and 1-flexion kernel (right) as a function of scaled radius for the polynomial indices used in the $K_{ap}$ reconstruction.  $Q$ and $r$ have been scaled by the aperture radius $R$ to make them unitless quantities.  A higher polynomial index implies sensitivity to a narrower range of radii within the aperture. The kernel is negative, meaning that 1-flexion measurements oriented towards the aperture center contribute positively to the aperture statistic.\label{fig:wqfunc}}
\end{figure}

We use the apertures $R=$45\arcsec, 60\arcsec, 75\arcsec, and 90\arcsec, and polynomial indices $l=$3, 5, and 7.  Figure \ref{fig:wqfunc} shows plots of $w(r)$ and $Q_{\Psi_1}(r)/R$ versus $r/R$ (so scaled to make them unitless quantities).  Higher polynomial indices make the kernel sensitive to a smaller range of radii within the aperture.  This implies that high-index kernels will detect smaller mass structures, though will also be more susceptible to noise fluctuations due to the finite number of flexion samplings within the aperture.  Lower indices reduce the amount of noise from discrete sampling, but also have a coarser resolution for mass structure and will therefore smooth out smaller structures.  As the aperture radius increases, the flexion signal correlated with the center of the aperture falls off quickly, again smoothing over small structures while large structures continue to be detected.

Because the systematics of this aperture statistic need to be further investigated, we evaluate the significance of the structure detected by constructing signal-to-noise ratio (SNR) maps from our flexion measurements instead of simply examining the $K_{ap}$ maps.  This is analogous to the use of ``B-mode'' convergence maps in other weak-lensing studies to evaluate the significance of mass structure detections, though it is more directly related to specific parameter uncertainty estimates.

\begin{figure}[h]
	\plottwo{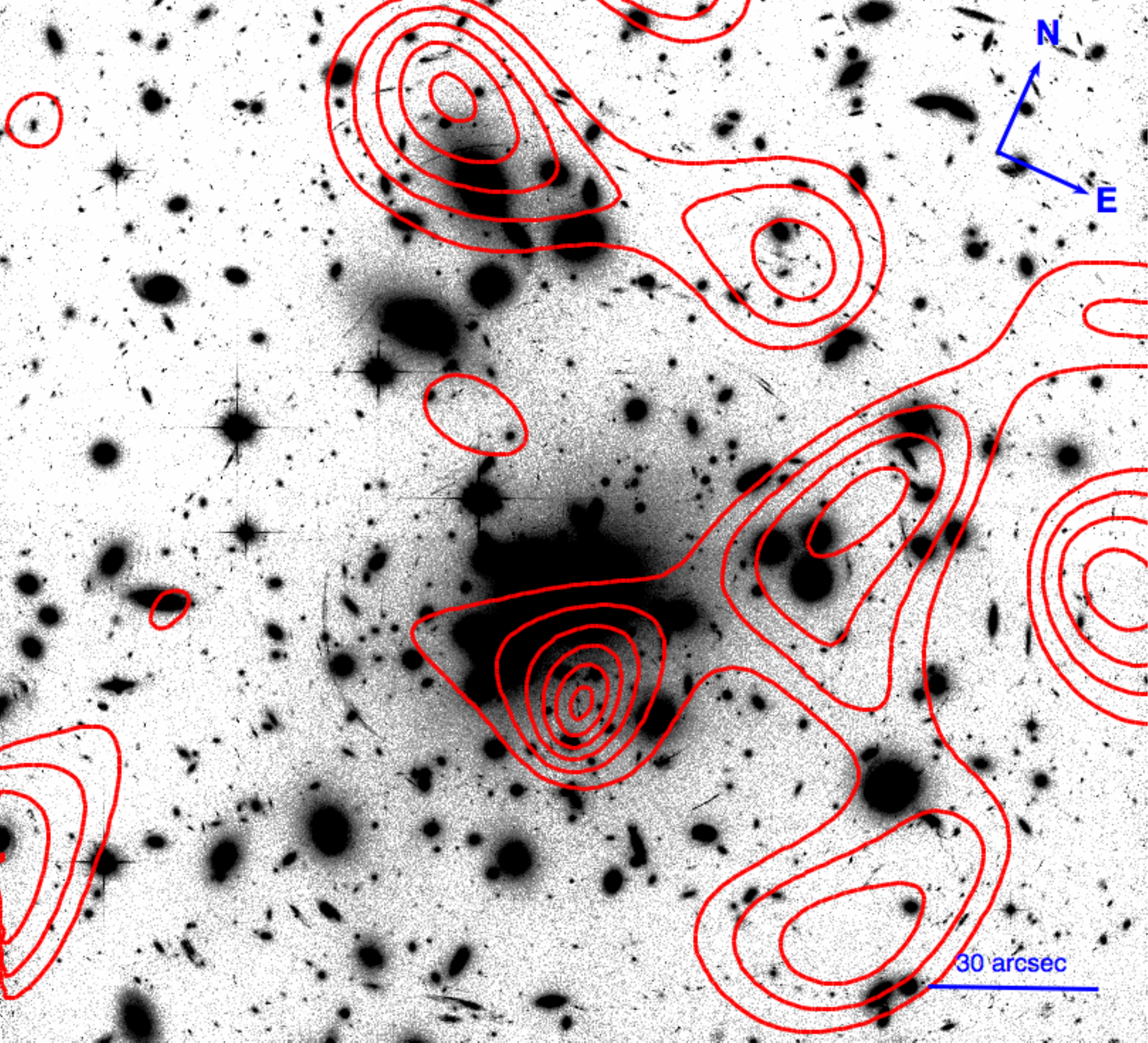}{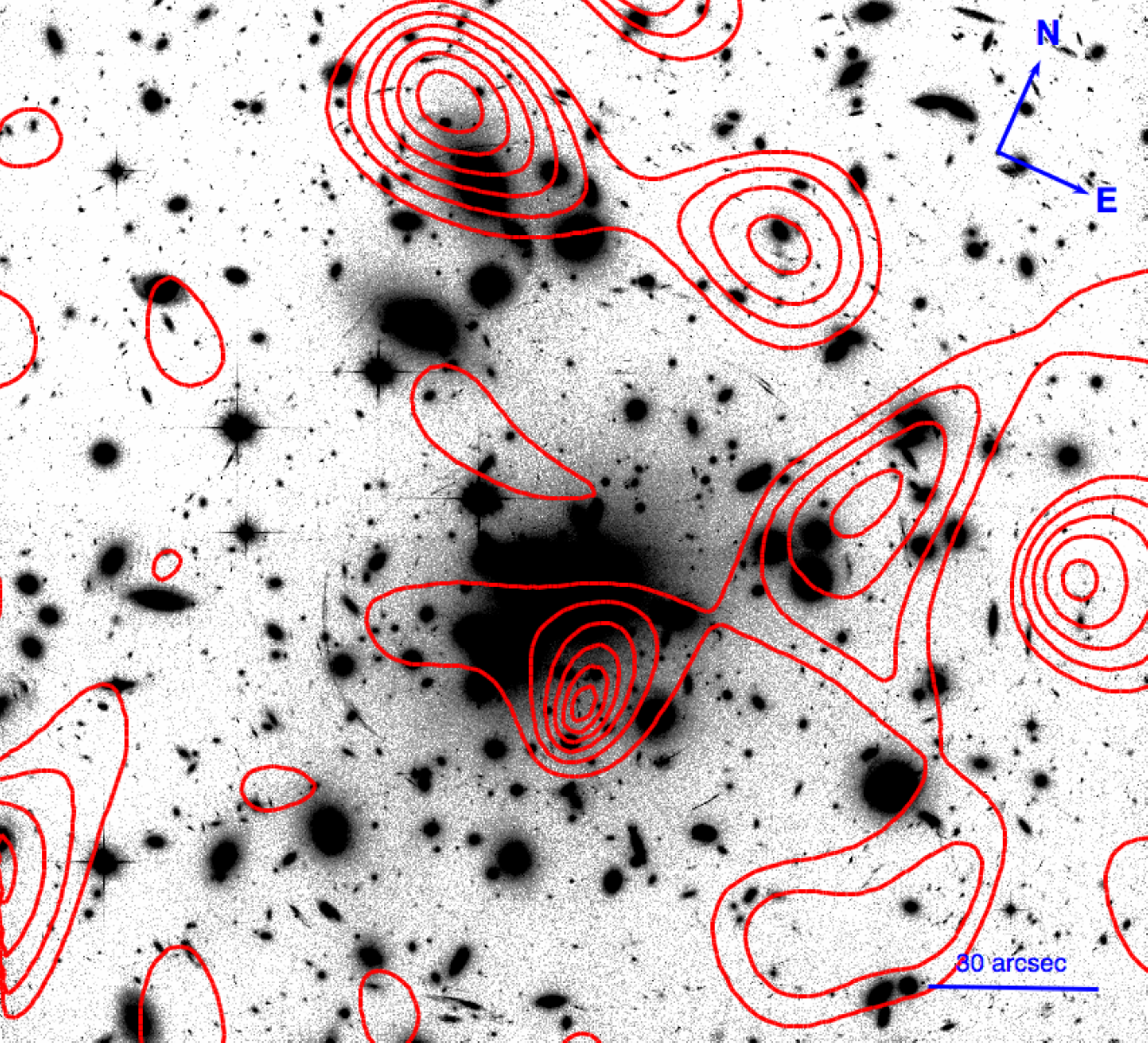}

	\plottwo{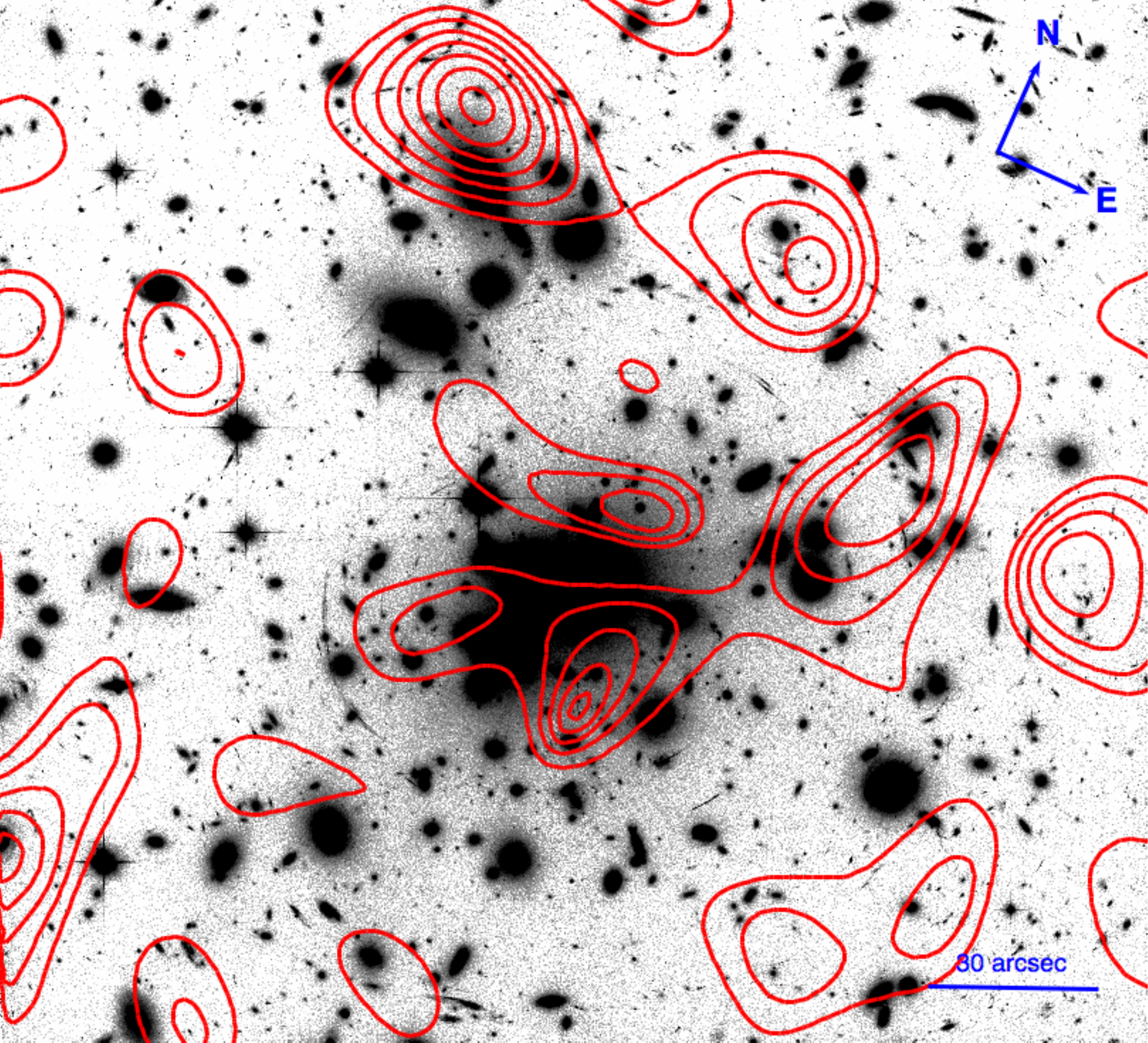}{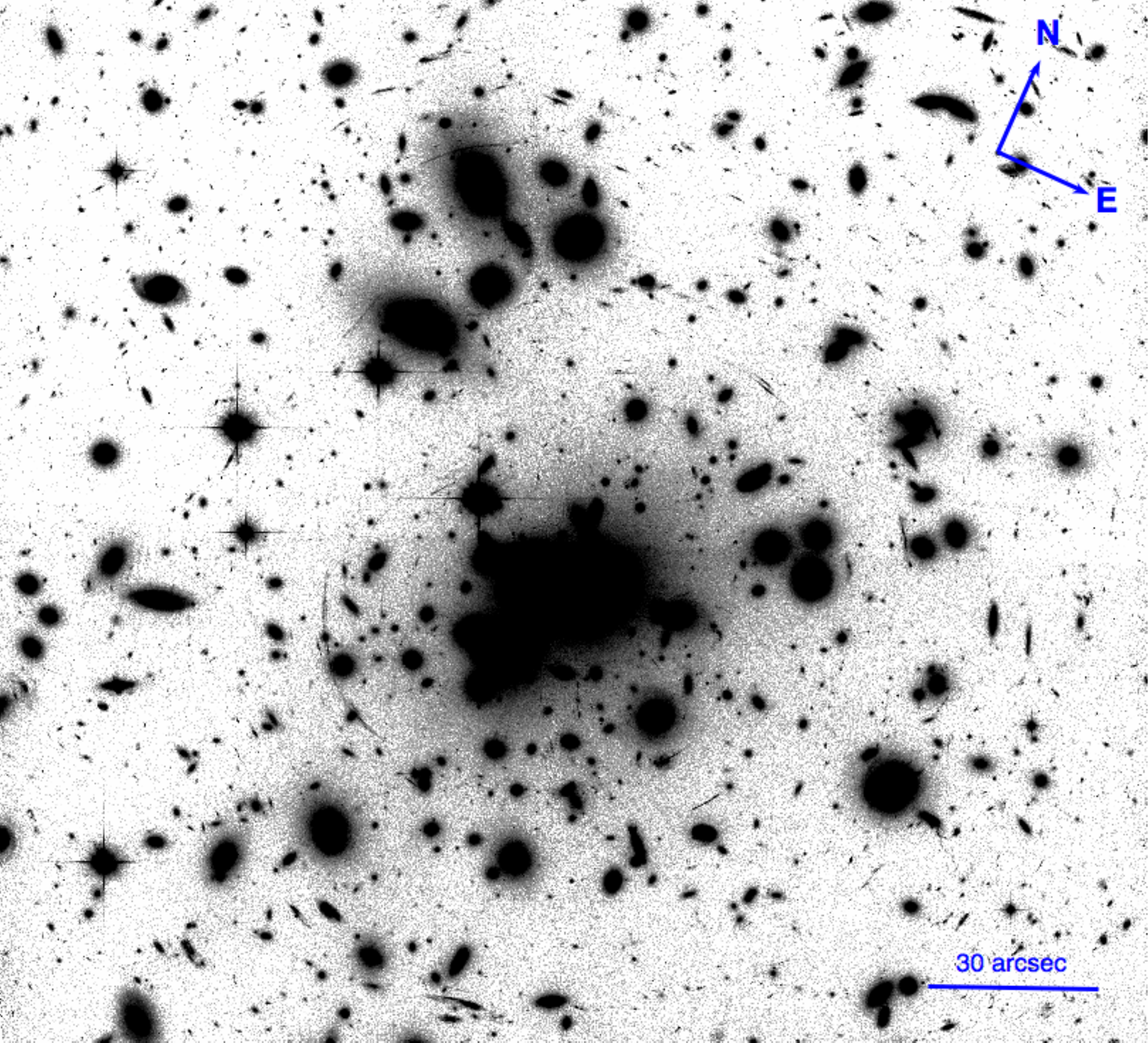}
	\caption{$K_{ap}$ signal-to-noise ratio (SNR) contours for $R=60\arcsec$ and three polynomial indices.  Top left: $l=3$; top right: $l=5$; bottom left: $l=7$; bottom right: four-filter coadded image without contours.  The SNR contours beginn at SNR=5 and increase in steps of 2.  Labels indicate celestial north/east and the angular scale. 1\arcsec=3.1 kpc at the cluster redshift.\label{fig:kap}}
\end{figure}

The procedure for producing the SNR maps is as follows:  For each combination of $R$ and $l$, we first construct a map of $K_{ap}$ on a 500$\times$500 grid of points $\theta_0$.  We then construct 1000 ``deviate'' maps.  For each deviate map, both components of the 1-flexion measurements for each object are shifted by a randomly-selected, normally-distributed value with standard deviation equal to the AIM fitting error for that flexion parameter.  For each grid point $\theta_0$, the standard deviation of the deviate $K_{ap}$ values at that grid point is used to calculate the SNR map value
\begin{equation}
	\text{SNR}(\theta_0)=\frac{K_{ap}(\theta_0)}{\sigma_{K_{ap}}(\theta_0)}.
\end{equation}
This yields maps of mass structure for each combination of $R$ and $l$.  Figure \ref{fig:kap} shows SNR contours overlaid on the F425W/F625W/F775W color image of A1689 for $R=60\arcsec$ and $l=$3, 5, and 7.  The major structures observable here are also detected in the other aperture sizes, though the larger apertures smooth over the smaller structures and reduce the magnitude of the SNR peaks.

The SNR maps do not represent quantitative statistical significance values, as we expect that there are some systematic effects included which are not yet quantified.  In particular, there are apparent edge effects from where the apertures extend beyond the observed field which produce spurious structures at the border of the observed field.  These edge effects seem to also skew some of the observed mass peaks towards the field edge.  However the persistence of structures despite varying apertures and polynomial indices, as well as their correlation with known mass structures (such as the visible subclustering of cluster member galaxies) indicate that these structures are sourced by actual physical structures.  In the context of these substructure reconstructions, a mass-sheet transformation $\kappa\rightarrow\lambda\kappa+1-\lambda$ is equivalent to $K\rightarrow K+\ln|\lambda|$, or simply a constant offset in $K$.

\section{Intrinsic Flexion}\label{sec:intrinsic}
An important consideration for including flexion into combined strong and weak lensing mass reconstructions in future work is quantifying the amount of flexion-like shape intrinsic to unlensed galaxy profiles.  Just as real galaxy profiles have a distribution of non-zero ellipticities which produce a scatter in the shear inferred from individual objects, there are also distributions of the third-order moments in unlensed galaxy profiles which create an ``intrinsic flexion'' signal.  This unlensed-shape-induced scatter is best characterized by the scatter in the dimensionless products $\alpha_{\text{obs}}\Psi_1$ and $\alpha_{\text{obs}}\Psi_3$, where $\alpha_{\text{obs}}$ is the observed, image plane size of the image.  The intrinsic flexion scatter should not be correlated with the astrophysically-sourced lensing flexion, so each component of the flexion fields can be used as an independent probe of the scatter along a single coordinate axis.

Table \ref{tab:intflex} lists the mean and standard deviation of these dimensionless products $\alpha_{\text{obs}}\Psi_{mn}$ for the flexion fields measured in A1689, characterized in three different object samples for each flexion field.  The full sample includes all objects used for the flexion analysis in \S\ref{sec:realdata}.  ``Low error'' and ``high error'' samples are defined independently for 1-flexion and 3-flexion based on the sum in quadrature of their respective flexion parameter error estimates: the full sample is divided about the median 1-flexion error of 0.029 arcsec$^{-1}$ and the median 3-flexion error of 0.049 arcsec$^{-1}$.  The scatter we observe is similar to values previously reported, though caution is warranted when directly comparing numerical values, due to the difference between our treatment of the mass-sheet degeneracy and that of \cite{goldberg+:07}.  

Another notable feature is the higher 1-flexion scatter in the low error subsample than in the high error sample, which we attribute to the astrophysical flexion signal.  The low error sample are those objects most likely to include a non-zero measured flexion value.  The objects which have larger true flexion field values are more likely to be well-constrained by the AIM fitting, and thus will more likely be in the low error sample and increase the scatter in our unitless figure-of-merit.  Sub-dividing the full sample by their radius from the cluster center has no discernible effect on the scatter, which is expected.  The flexion signal is less dependent on the overall cluster potential, and more dependent on the local substructure.  A detailed study of the scatter in the measured flexion fields and the relative contributions from galaxy shape, galaxy-galaxy lensing, and cluster substructure lensing,  is beyond the scope of this paper, but will be addressed in future work.  3-flexion measurements are typically noisier than 1-flexion, and the increase in scatter in the high noise sample is an indication that the 3-flexion scatter may be dominated by the measurement noise.

\section{Summary and Discussion}\label{sec:summary}
Obtaining detailed measurements of the mass structure in galaxy clusters is an important step in understanding these important cosmological objects.  To this end, a variety of mass reconstruction techniques based on measuring the lensing distortion of background galaxies have been developed.  The combination of multiple orders of lensing effects, such as weak lensing shear with strong lensing measurements, allows for a mass reconstruction over a large area and a range of lensing field strengths.  

However, near the center of the cluster the convergence can vary significantly over the scale on which galaxy ellipticities are typically averaged to extract the weak-lensing shear, increasing the scatter in the shear measurements and reducing the resolution of the weak lensing mass maps in these areas.  It is not, however, possible to measure a strong lensing signal for all background galaxies in this regime since not all objects will be multiply imaged.  An alternative approach is necessary to extract the lensing information from these distorted background images.  Second-order weak lensing distortions, called flexion, are high signal-to-noise probes of mass substructure on small angular scales.  Flexion measurements alone can be used to measure mass structure, and including them in joint weak+strong lensing measurements of cluster mass distributions will better resolve cluster substructures.

In recent years, two main techniques for measuring flexion (shapelets and HOLICs) have been applied to measure mass structure in the galaxy cluster Abell 1689.  We have developed an alternative method, the Analytic Image Model (AIM) method.  This approach has several desirable properties for flexion measurement.  Because the AIM method is based on model optimization, a full covariance matrix including both lensing and non-lensing parameters accompanies the flexion field estimates for each observed background galaxy.  This provides direct error estimates for all components of the flexion fields as well as an indicator of parameter degeneracies.  Additionally, the AIM method is fully invariant to the mass-sheet degeneracy, a property that previous flexion analyses of observational data have not had.  And because the AIM method forward-models quadratic lensing rather than inverting measured properties to obtain flexion estimates, it provides an important systematic check on the shaplet and HOLIC based methods.

Using simulated data images across a broad span of parameter space, we have demonstrated that the AIM method accurately reproduces both the lensing parameters and the intrinsic shape parameters.  By fixing the shear parameters to an estimate of the true shear field, we recover the flexion parameters.  We also recover the intrinsic shape parameters up to the error in the ellipticity induced by an erroneous fixed shear value.  These simulations validate the accuracy of the AIM method for measuring flexion.

We applied the AIM method to measure flexion fields in the galaxy cluster Abell 1689, and used a modified version of the 1-flexion aperture mass statistic presented by \cite{leonard+:10}, accounting for the mass-sheet degeneracy, to infer mass structures which are consistent with previous measurements.  The structures we observe are significantly detected using multiple aperture functions, meaning that they are sourced by physical structures in the cluster.  This application both highlights the utility of flexion measurements for mass reconstructions in clusters, as well as the validity of the flexion measurements obtained via the AIM method.

A robust literature of weak lensing shear and strong lensing mass measurements already exists, and many studies combine both weak and strong lensing to improve constraints on the mass distribution in clusters of galaxies.  Including flexion in these multi-scale mass reconstructions will allow for a more detailed mapping of the mass distribution in the central parts of clusters, regions where the linear weak lensing approximation does not capture all of the measurable lensing information and strong lensing data is not ubiquitous.  A complete comparison between the AIM method and the two previously described methods remains to be done; such a comparison will be able to characterize the noise properties and systematics of each technique, as well as their relative strengths and weaknesses.  Characterization of the noise from flexion-like shape properties in unlensed galaxy images and an understanding of how that noise contaminates the lensing flexion signal is needed for accurate mass reconstructions.  Due to its fundamentally different approach, the AIM method will provide an important constraint on the intrinsic noise of flexion measurements, separating out the systematic effects introduced by measurement techniques.

\acknowledgments
BC would like to thank Maru\v{s}a Brada\v{c} for helpful discussions and feedback on the manuscript.  Support for this work was provided by NASA through SAO Award Number 2834-MIT-SAO-4018 issued by the Chandra X-Ray Observatory Center, which is operated by the Smithsonian Astrophysical Observatory for and on behalf of NASA under contract NAS8-03060, and through grant HST-GO-11099 from the Space Telescope Science Institute (STScI), which is operated by AURA, Inc., under NASA contract NAS5-26555 and NNX08AD79G.

Results in this paper were based on observations made with the NASA/ESA Hubble Space Telescope, and obtained from the Hubble Legacy Archive, which is a collaboration between the Space Telescope Science Institute (STScI/NASA), the Space Telescope European Coordinating Facility (ST-ECF/ESA) and the Canadian Astronomy Data Centre (CADC/NRC/CSA).

\newpage

\begin{deluxetable}{clc}
\tablecolumns{4}
\tablewidth{0pt} 
\tablecaption{Allowed ranges for model fit parameters.  $L$ is the full side length of the single-galaxy data image being fit, which is chosen to be 1.5 times the observed semi-major axis length.  Angular units assume a HST ACS WFC pixel scale of 0\farcs05/pixel, as in \S\ref{sec:realdata}.\label{tab:pars}}
\tablehead{
	\colhead{Parameter}		&
	\colhead{Significance}	&
	\colhead{Range}			}
\startdata
	$\Psi_{11}$, $\Psi_{12}$		& Reduced 1-flexion	& $-2$ to 2 arcsec$^{-1}$	\\
	$\Psi_{31}$, $\Psi_{32}$		& Reduced 3-flexion	& $-2$ to 2 arcsec$^{-1}$	\\
	\tableline
	$\log S_0$					& Total flux		&	$-10$ to 10					\\
	$\alpha$						& Image Size		&	0.05 to $3L$ arcsec			\\
	$\theta_{c1}$, $\theta_{c2}$	& Image Center	&	$-L$ to $L$ arcsec			\\
	$\epsilon_1$, $\epsilon_2$	& Ellipticity	&	$-1$ to 1					\\
\enddata
\end{deluxetable}

\begin{deluxetable}{lcccc}
\tablecolumns{5}
\tablewidth{0pt} 
\tablecaption{Mean and standard deviations for the products using the SExtractor-determined image sizes in the full, low error, and high error object samples.\label{tab:intflex}}
\tablehead{
	\colhead{Sample}								&
	\colhead{$\langle\alpha\Psi_{1n}\rangle$}		&
	\colhead{$\sigma_{\alpha\Psi_{1n}}$}			&
	\colhead{$\langle\alpha\Psi_{3n}\rangle$}		&
	\colhead{$\sigma_{\alpha\Psi_{3n}}$}			}
\startdata
	Full				&	0.000	&	0.042	&	0.000	&	0.048	\\
	Low Error		&	0.004	&	0.047	&	0.000	&	0.041	\\
	High Error		&	-0.004	&	0.037	&	-0.001	&	0.055	\\
\enddata
\end{deluxetable}

\begin{deluxetable}{c|cc|cc|cc|cc|cc|cc|cc|cc|cc|cc|cc}
\rotate
\tablecolumns{23}
\tabletypesize{\scriptsize}
\setlength{\tabcolsep}{0.025in} 
\tablewidth{0pt}
\tablecaption{Position, fit, and error values for the model parameters of the 50 objects in the A1689 field with the best 1-flexion estimates.  The RA and Dec coordinates are J2000 epoch.  $S_0$ has units of $e^-$/s.  $\theta_{cn}$ and $\alpha$ have units of arcseconds.  Each flexion parameter $\Psi_{mn}$ has units of arcsec$^{-1}$. The ellipticity parameters $\epsilon_n$ are unitless.\label{tab:fitresults}}
\tablehead{
	\colhead{Object}					&
	\colhead{RA}						&
	\colhead{Dec}					&
	\colhead{$\log S_0$}				&
	\colhead{$\sigma_{\log S_0}$}	&
	\colhead{$\theta_{c1}$}			&
	\colhead{$\sigma_{\theta_{c1}}$}	&
	\colhead{$\theta_{c2}$}			&
	\colhead{$\sigma_{\theta_{c2}}$}	&
	\colhead{$\alpha$}				&
	\colhead{$\sigma_{\alpha}$}		&
	\colhead{$\epsilon_1$}			&
	\colhead{$\sigma_{\epsilon_1}$}	&
	\colhead{$\epsilon_2$}			&
	\colhead{$\sigma_{\epsilon_2}$}	&
	\colhead{$\Psi_{11}$}			&
	\colhead{$\sigma_{\Psi_{11}}$}	&
	\colhead{$\Psi_{12}$}			&
	\colhead{$\sigma_{\Psi_{12}}$}	&
	\colhead{$\Psi_{31}$}			&
	\colhead{$\sigma_{\Psi_{31}}$}	&
	\colhead{$\Psi_{32}$}			&
	\colhead{$\sigma_{\Psi_{32}}$}	}
\startdata
1 & 13:11:34.5 & -1:19:32.1 & 0.077 & 0.050 & -0.049 & 0.002 & -0.086 & 0.000 & 0.148 & 0.002 & -0.187 & 0.009 & 0.541 & 0.009 & 0.268 & 0.007 & -0.060 & 0.008 & -0.063 & 0.022 & -0.278 & 0.014 \\ 
2 & 13:11:28.5 & -1:20:60.0 & -0.104 & 0.012 & 0.009 & 0.003 & -0.076 & 0.005 & 0.135 & 0.003 & -0.235 & 0.017 & 0.027 & 0.018 & -0.071 & 0.010 & 0.345 & 0.007 & 0.068 & 0.019 & 0.316 & 0.016 \\ 
3 & 13:11:28.8 & -1:20:13.5 & -0.081 & 0.005 & -0.074 & 0.005 & -0.046 & 0.004 & 0.114 & 0.002 & 0.231 & 0.000 & 0.478 & 0.016 & 0.224 & 0.009 & 0.091 & 0.009 & -0.051 & 0.005 & -0.353 & 0.028 \\ 
4 & 13:11:22.2 & -1:21:00.5 & 0.553 & 0.050 & -0.038 & 0.005 & -0.033 & 0.007 & 0.283 & 0.002 & -0.121 & 0.007 & 0.291 & 0.007 & 0.065 & 0.010 & -0.027 & 0.010 & 0.042 & 0.015 & -0.092 & 0.015 \\ 
5 & 13:11:31.5 & -1:20:37.1 & -0.301 & 0.015 & -0.011 & 0.004 & -0.128 & 0.006 & 0.121 & 0.003 & -0.261 & 0.022 & -0.178 & 0.019 & 0.102 & 0.014 & 0.437 & 0.006 & 0.311 & 0.018 & -0.031 & 0.013 \\ 
6 & 13:11:27.8 & -1:20:32.3 & 0.482 & 0.006 & -0.042 & 0.006 & -0.027 & 0.006 & 0.267 & 0.003 & 0.021 & 0.009 & 0.034 & 0.009 & 0.054 & 0.011 & -0.012 & 0.012 & -0.045 & 0.019 & 0.042 & 0.019 \\ 
7 & 13:11:26.8 & -1:19:05.0 & 0.409 & 0.050 & -0.056 & 0.005 & -0.032 & 0.006 & 0.265 & 0.002 & -0.069 & 0.009 & 0.005 & 0.009 & 0.012 & 0.012 & 0.030 & 0.011 & 0.020 & 0.019 & 0.030 & 0.019 \\ 
8 & 13:11:29.9 & -1:19:01.4 & -0.424 & 0.005 & 0.062 & 0.000 & -0.025 & 0.000 & 0.114 & 0.004 & 0.531 & 0.023 & 0.063 & 0.000 & -0.164 & 0.010 & 0.275 & 0.013 & -0.228 & 0.054 & -0.275 & 0.017 \\ 
9 & 13:11:22.5 & -1:21:12.8 & 0.159 & 0.009 & 0.008 & 0.006 & 0.055 & 0.004 & 0.182 & 0.003 & 0.155 & 0.012 & 0.223 & 0.013 & -0.016 & 0.014 & -0.229 & 0.010 & -0.106 & 0.021 & -0.013 & 0.023 \\ 
10 & 13:11:34.4 & -1:21:32.3 & -0.085 & 0.005 & -0.068 & 0.008 & -0.017 & 0.003 & 0.179 & 0.004 & 0.417 & 0.000 & -0.368 & 0.015 & -0.014 & 0.002 & -0.130 & 0.017 & 0.212 & 0.040 & -0.058 & 0.009 \\ 
11 & 13:11:28.6 & -1:20:31.8 & -0.393 & 0.022 & 0.031 & 0.006 & -0.107 & 0.007 & 0.104 & 0.004 & -0.091 & 0.037 & -0.121 & 0.040 & -0.302 & 0.012 & 0.258 & 0.012 & -0.260 & 0.027 & -0.179 & 0.026 \\ 
12 & 13:11:25.3 & -1:19:57.7 & -0.295 & 0.015 & 0.006 & 0.005 & -0.050 & 0.006 & 0.133 & 0.004 & -0.138 & 0.023 & -0.288 & 0.024 & -0.354 & 0.010 & 0.043 & 0.015 & 0.191 & 0.032 & -0.174 & 0.023 \\ 
13 & 13:11:22.5 & -1:20:20.7 & 0.499 & 0.008 & -0.026 & 0.009 & -0.099 & 0.006 & 0.272 & 0.004 & 0.157 & 0.012 & -0.112 & 0.011 & 0.015 & 0.014 & 0.107 & 0.011 & 0.001 & 0.021 & -0.129 & 0.021 \\ 
14 & 13:11:35.8 & -1:20:59.6 & -0.194 & 0.015 & 0.081 & 0.006 & -0.174 & 0.007 & 0.136 & 0.004 & -0.071 & 0.018 & -0.506 & 0.019 & 0.026 & 0.012 & 0.279 & 0.014 & -0.051 & 0.024 & 0.154 & 0.024 \\ 
15 & 13:11:32.4 & -1:20:37.2 & -0.183 & 0.013 & -0.032 & 0.005 & 0.022 & 0.005 & 0.146 & 0.003 & 0.041 & 0.022 & 0.100 & 0.022 & 0.131 & 0.014 & -0.294 & 0.011 & 0.176 & 0.027 & -0.091 & 0.030 \\ 
16 & 13:11:22.6 & -1:21:02.1 & -0.046 & 0.022 & -0.054 & 0.015 & -0.129 & 0.016 & 0.220 & 0.007 & -0.051 & 0.021 & 0.482 & 0.026 & -0.129 & 0.012 & 0.281 & 0.014 & -0.069 & 0.023 & 0.309 & 0.023 \\ 
17 & 13:11:38.4 & -1:19:33.0 & 0.049 & 0.050 & 0.012 & 0.004 & -0.053 & 0.002 & 0.136 & 0.002 & 0.179 & 0.012 & 0.032 & 0.012 & -0.188 & 0.014 & 0.152 & 0.012 & 0.059 & 0.030 & -0.275 & 0.028 \\ 
18 & 13:11:34.6 & -1:19:45.9 & 0.156 & 0.050 & 0.000 & 0.004 & 0.006 & 0.004 & 0.169 & 0.002 & 0.034 & 0.010 & -0.271 & 0.011 & -0.110 & 0.012 & -0.126 & 0.015 & 0.126 & 0.025 & -0.295 & 0.026 \\ 
19 & 13:11:26.6 & -1:20:08.5 & -0.072 & 0.013 & -0.113 & 0.005 & 0.023 & 0.006 & 0.172 & 0.004 & -0.051 & 0.020 & -0.011 & 0.021 & 0.248 & 0.011 & -0.069 & 0.015 & -0.078 & 0.028 & 0.178 & 0.028 \\ 
20 & 13:11:34.3 & -1:21:47.7 & 0.229 & 0.011 & -0.077 & 0.010 & -0.077 & 0.012 & 0.253 & 0.005 & -0.096 & 0.016 & -0.382 & 0.015 & 0.141 & 0.012 & 0.131 & 0.016 & 0.102 & 0.025 & 0.126 & 0.022 \\ 
21 & 13:11:27.8 & -1:20:34.3 & 0.221 & 0.050 & -0.025 & 0.002 & 0.011 & 0.002 & 0.118 & 0.001 & -0.036 & 0.009 & -0.168 & 0.008 & -0.186 & 0.016 & -0.170 & 0.014 & 0.078 & 0.029 & -0.049 & 0.029 \\ 
22 & 13:11:26.1 & -1:19:43.0 & -0.234 & 0.050 & 0.013 & 0.003 & -0.087 & 0.004 & 0.104 & 0.001 & -0.168 & 0.016 & -0.440 & 0.015 & -0.279 & 0.018 & 0.024 & 0.012 & 0.231 & 0.059 & -0.117 & 0.034 \\ 
23 & 13:11:24.1 & -1:19:49.7 & -0.029 & 0.050 & 0.000 & 0.003 & -0.032 & 0.009 & 0.148 & 0.002 & -0.472 & 0.011 & -0.265 & 0.012 & -0.249 & 0.014 & -0.071 & 0.017 & -0.063 & 0.028 & -0.292 & 0.028 \\ 
24 & 13:11:28.8 & -1:20:06.6 & 0.384 & 0.008 & 0.009 & 0.009 & -0.073 & 0.008 & 0.282 & 0.004 & 0.065 & 0.012 & 0.076 & 0.013 & -0.020 & 0.015 & 0.055 & 0.016 & -0.043 & 0.025 & 0.078 & 0.025 \\ 
25 & 13:11:21.9 & -1:21:09.7 & 0.273 & 0.008 & -0.002 & 0.007 & -0.007 & 0.005 & 0.222 & 0.003 & 0.104 & 0.012 & -0.039 & 0.013 & -0.108 & 0.014 & -0.038 & 0.017 & 0.048 & 0.026 & -0.088 & 0.026 \\
26 & 13:11:30.5 & -1:22:10.8 & 0.216 & 0.021 & -0.278 & 0.016 & 0.100 & 0.010 & 0.233 & 0.008 & 0.308 & 0.018 & -0.516 & 0.018 & 0.144 & 0.017 & 0.212 & 0.013 & -0.185 & 0.016 & -0.109 & 0.025 \\ 
27 & 13:11:29.9 & -1:19:51.5 & -0.242 & 0.015 & -0.090 & 0.005 & -0.049 & 0.007 & 0.152 & 0.004 & -0.126 & 0.025 & -0.043 & 0.025 & 0.286 & 0.012 & 0.111 & 0.018 & -0.073 & 0.034 & -0.123 & 0.030 \\ 
28 & 13:11:29.2 & -1:19:32.3 & 0.275 & 0.050 & 0.004 & 0.003 & -0.070 & 0.004 & 0.175 & 0.002 & -0.200 & 0.008 & -0.162 & 0.008 & -0.014 & 0.018 & 0.079 & 0.013 & 0.001 & 0.025 & 0.047 & 0.025 \\ 
29 & 13:11:30.1 & -1:21:06.5 & -0.033 & 0.013 & -0.142 & 0.006 & -0.034 & 0.005 & 0.166 & 0.004 & 0.094 & 0.019 & 0.144 & 0.020 & 0.222 & 0.014 & -0.015 & 0.018 & -0.201 & 0.028 & -0.216 & 0.036 \\ 
30 & 13:11:33.6 & -1:21:06.7 & -0.383 & 0.013 & -0.024 & 0.004 & -0.113 & 0.005 & 0.101 & 0.003 & -0.153 & 0.023 & -0.254 & 0.023 & 0.219 & 0.015 & 0.272 & 0.017 & 0.270 & 0.039 & -0.057 & 0.050 \\ 
31 & 13:11:28.3 & -1:20:03.9 & -0.401 & 0.024 & -0.072 & 0.008 & 0.078 & 0.010 & 0.147 & 0.006 & -0.107 & 0.037 & -0.168 & 0.036 & 0.243 & 0.014 & -0.219 & 0.019 & 0.324 & 0.038 & 0.105 & 0.033 \\ 
32 & 13:11:33.5 & -1:19:11.4 & -0.293 & 0.025 & 0.115 & 0.014 & 0.014 & 0.008 & 0.160 & 0.007 & 0.346 & 0.026 & 0.385 & 0.026 & -0.297 & 0.012 & 0.089 & 0.020 & 0.198 & 0.022 & -0.025 & 0.035 \\ 
33 & 13:11:25.4 & -1:21:16.2 & -0.159 & 0.026 & -0.091 & 0.010 & -0.023 & 0.010 & 0.204 & 0.008 & -0.013 & 0.037 & 0.071 & 0.028 & 0.404 & 0.012 & -0.075 & 0.021 & 0.228 & 0.028 & 0.066 & 0.031 \\ 
34 & 13:11:28.6 & -1:19:11.3 & -0.134 & 0.050 & 0.005 & 0.003 & -0.064 & 0.003 & 0.119 & 0.002 & -0.054 & 0.015 & -0.093 & 0.015 & -0.014 & 0.023 & 0.312 & 0.008 & -0.059 & 0.032 & 0.096 & 0.025 \\ 
35 & 13:11:34.4 & -1:21:11.0 & 0.523 & 0.010 & -0.069 & 0.010 & -0.033 & 0.017 & 0.386 & 0.006 & -0.180 & 0.014 & 0.028 & 0.014 & 0.047 & 0.015 & 0.001 & 0.019 & -0.017 & 0.023 & 0.009 & 0.024 \\ 
36 & 13:11:24.2 & -1:20:33.9 & -0.479 & 0.019 & -0.048 & 0.006 & 0.053 & 0.007 & 0.118 & 0.004 & -0.093 & 0.031 & 0.234 & 0.032 & 0.125 & 0.021 & -0.332 & 0.013 & 0.177 & 0.041 & -0.140 & 0.037 \\ 
37 & 13:11:35.1 & -1:19:13.4 & 0.164 & 0.050 & 0.036 & 0.004 & -0.015 & 0.003 & 0.158 & 0.002 & 0.120 & 0.010 & 0.206 & 0.010 & -0.159 & 0.013 & 0.015 & 0.021 & -0.046 & 0.030 & 0.073 & 0.028 \\ 
38 & 13:11:26.5 & -1:20:56.2 & 0.028 & 0.010 & -0.071 & 0.006 & -0.017 & 0.004 & 0.163 & 0.003 & 0.272 & 0.013 & 0.236 & 0.014 & 0.183 & 0.013 & -0.102 & 0.021 & -0.178 & 0.027 & -0.079 & 0.033 \\ 
39 & 13:11:30.5 & -1:22:10.5 & 0.106 & 0.029 & 0.021 & 0.016 & -0.080 & 0.009 & 0.205 & 0.008 & 0.256 & 0.020 & -0.420 & 0.027 & 0.346 & 0.019 & 0.025 & 0.016 & -0.203 & 0.022 & -0.352 & 0.040 \\ 
40 & 13:11:27.1 & -1:18:39.9 & -0.133 & 0.050 & -0.069 & 0.004 & -0.040 & 0.004 & 0.118 & 0.002 & 0.002 & 0.015 & -0.295 & 0.014 & 0.194 & 0.017 & 0.227 & 0.018 & 0.073 & 0.041 & 0.003 & 0.038 \\ 
41 & 13:11:22.7 & -1:21:01.3 & -0.192 & 0.019 & -0.066 & 0.010 & -0.066 & 0.019 & 0.206 & 0.006 & -0.295 & 0.025 & -0.305 & 0.027 & 0.353 & 0.013 & 0.141 & 0.021 & -0.021 & 0.029 & 0.178 & 0.029 \\ 
42 & 13:11:28.2 & -1:20:07.3 & -0.103 & 0.019 & -0.088 & 0.010 & -0.040 & 0.011 & 0.197 & 0.006 & -0.028 & 0.027 & 0.258 & 0.028 & 0.275 & 0.013 & -0.089 & 0.022 & -0.070 & 0.032 & -0.009 & 0.033 \\ 
43 & 13:11:36.9 & -1:19:45.6 & 0.026 & 0.010 & 0.002 & 0.004 & 0.049 & 0.005 & 0.167 & 0.003 & -0.166 & 0.014 & 0.011 & 0.016 & -0.009 & 0.021 & -0.186 & 0.014 & -0.006 & 0.035 & -0.185 & 0.031 \\ 
44 & 13:11:25.6 & -1:19:49.5 & 0.274 & 0.050 & -0.015 & 0.003 & -0.065 & 0.007 & 0.199 & 0.002 & -0.341 & 0.009 & -0.048 & 0.008 & -0.005 & 0.022 & 0.026 & 0.013 & 0.009 & 0.027 & -0.004 & 0.025 \\ 
45 & 13:11:36.5 & -1:19:19.2 & 0.109 & 0.050 & -0.013 & 0.004 & 0.026 & 0.004 & 0.157 & 0.002 & -0.060 & 0.012 & 0.082 & 0.012 & -0.059 & 0.020 & -0.164 & 0.016 & 0.065 & 0.032 & -0.106 & 0.031 \\ 
46 & 13:11:30.3 & -1:18:37.0 & 0.214 & 0.008 & 0.009 & 0.008 & -0.045 & 0.008 & 0.242 & 0.003 & -0.003 & 0.013 & 0.006 & 0.013 & -0.017 & 0.018 & -0.003 & 0.018 & -0.074 & 0.031 & 0.044 & 0.031 \\ 
47 & 13:11:26.2 & -1:21:19.9 & 0.134 & 0.011 & -0.083 & 0.009 & -0.053 & 0.007 & 0.232 & 0.004 & 0.107 & 0.016 & 0.062 & 0.018 & 0.108 & 0.017 & 0.017 & 0.019 & -0.143 & 0.032 & -0.084 & 0.032 \\ 
48 & 13:11:32.7 & -1:21:08.7 & 0.177 & 0.050 & 0.010 & 0.004 & 0.007 & 0.004 & 0.164 & 0.002 & -0.027 & 0.010 & 0.075 & 0.011 & -0.098 & 0.019 & -0.071 & 0.017 & 0.119 & 0.032 & -0.049 & 0.032 \\ 
49 & 13:11:36.8 & -1:20:36.8 & -0.313 & 0.014 & 0.011 & 0.004 & -0.054 & 0.005 & 0.118 & 0.003 & -0.061 & 0.025 & 0.022 & 0.024 & -0.294 & 0.018 & 0.177 & 0.019 & 0.041 & 0.041 & -0.207 & 0.034 \\ 
50 & 13:11:27.3 & -1:19:50.0 & 0.063 & 0.011 & -0.021 & 0.007 & -0.072 & 0.005 & 0.185 & 0.003 & 0.104 & 0.016 & 0.061 & 0.017 & -0.014 & 0.019 & 0.173 & 0.018 & 0.006 & 0.035 & 0.063 & 0.035 \\ 
\enddata
\end{deluxetable}

\clearpage

\end{document}